\documentclass{nseJournal}
\usepackage{longtable}
\makeatletter
\def\@authors{\@empty}

\renewcommand\addAuthor[2]{%
  \ifx\@authors\@empty
    \gdef\@authors{#1$^{\text{#2}}$}%
  \else
    \g@addto@macro\@authors{, #1$^{\text{#2}}$}%
  \fi
}

\def\@affiliations{\@empty}
\renewcommand\addAffiliation[2]{%
  \g@addto@macro\@affiliations{\tabularnewline\tabularnewline$^{#1}$#2}%
}
\makeatother

\begin{document}
%\linenumbers
\title{Design and Commissioning of a Deuterium-Tritium Gas Delivery System for Muon Catalyzed Fusion in a Diamond Anvil Cell}

%-------------------------
% Authors and affiliations
%-------------------------
% Mapping:
% 1 -> a  Acceleron Fusion, USA
% 2 -> b  Torion USA, USA
% 3 -> c  Torion Plasma, USA
% 4 -> d  Tritium Solutions, Inc., USA
% 5 -> e  Paul Scherrer Institute (PSI), Switzerland
% 6 -> f  ETH Zürich, Switzerland
% 7 -> g  NK Labs, USA
% 8 -> h   Oxford University, Oxford, England
% 9 -> i   Laboratory for Laser Energetics, University of Rochester, Rochester, NY, USA
% 10 -> j  Institute for Particle Physics and Astrophysics, ETH Zurich, Zurich, Switzerland
% 11 -> k  Massachusetts Institute of Technology, Cambridge, MA, USA
% 12 -> l  REB Research and Consulting, Oak Park, MI, USA
\addAuthor{E.~Koukina}{a} % First author - Wrote paper, led gas delivery system project, designed DD system, contributions to design of DT system, substantial experimental operations     

% People who did substantial work on gas delivery system, including control system, raman spec, tritium safety, tritum compatibility of DAC, controlling the flow of tritium during the experiment, tritium safety, renders of CAD 
\addAuthor{C.~Fagan}{b}   % Designed and built the DT system, substantial experimental operations % comments incorprated 
       
\addAuthor{C.~R.~Shmayda}{c} % Designed the VAC-SEC and electronics for the DT system, piping, operating system
% approved publishing
\addAuthor{J.~D.~Kalow}{a}  % Designed the anvil cell and minichamber described in the paper, developed and carried out DAC-side tritium loading procedure for tritum, designed and built the vacuum pump train and stack connection manifolds for the DD and DT system
\addAuthor{D.~M.~Harrington}{a} % Designed the control system for the DD system, designed the optics system used to collect the raman plot shown, contributions to the FMEA
\addAuthor{G.~Harris}{a} % Substantial experimental operations, development of gas delivery piping in the vacuum chamber, developed the brazing process to remove polymers from the cell, process development for tritium-compatible minichambers
\addAuthor{K.~McCormack}{a} % Substantial experimental operations of DD system in 2025, development and carrying out of bakeout procedures, writing for future work section
\addAuthor{M.~Mundt}{f} % Collected and analyzed data from the Raman spectrometer for plot shown in the paper, writing for Raman spectrometer section
\addAuthor{K.~Lau}{f}  % Wrote the PLC sofware for the DD and DT control system
\addAuthor{D.~Zajac}{f} % Wrote the ROS software for the DD and DT control systems
\addAuthor{M.~Koch}{b} % Substantial experimental operations and piping, operation of the VAC-SEC and DT system
% comments incorprated 
\addAuthor{S.~Varner}{a} % Created the CAD renderings in the paper, substantial contributions to overall CAD design of the system pictured in the paper
% Comments in overleaf, approved publishing
\addAuthor{A.~Golossanov}{a,e}  % Substantial experimental operations related to tritium safety, fume hood, stack piping, regulatory approval.
\addAuthor{S.~Bull}{a} % Substantial experimental operations, building controllers, piping, developement of wiring/wrapping approach for bakeout heaters
\addAuthor{R.~Buxbaum}{l} % Built DD system permeator, consultation on gas purity and system design 
% approved publishing
\addAuthor{W.~Stadolnik}{f} % Contributions to control system design and bakout subsystem

% Contributions to overall experiment design as desribed in the introduction, not specific to gas delivery - alphabetical order by last name
\addAuthor{J.~A.~Allen}{f}         
\addAuthor{J.~Betances}{f}
\addAuthor{N.~J.~Brennan}{f} 
\addAuthor{R.~Chaney}{f} % approved
\addAuthor{W.~R.~Cutler}{a,h} % William Reuel Cutler 
\addAuthor{J.~Davies}{i} % pending LLE review
\addAuthor{C.~Forrest}{i}
\addAuthor{P.~Ga\-ndhi}{g}
\addAuthor{J.~T.~Hinchen}{f} % approved
\addAuthor{C.~J.~Johnstone}{g}
\addAuthor{K.~Kem}{f} %approved
\addAuthor{M.~Khan\-daker}{a}
\addAuthor{M.~Kiburg}{g} % approved pending other fnal approvals
\addAuthor{I.~Kiniti}{f} % approved
\addAuthor{A.~D.~Knaian}{f}
\addAuthor{L.~E.~Knaian}{a}
\addAuthor{N.~J.~L.~MacFadden}{f,m} % Nate James Lewkowicz MacFadden 
\addAuthor{D.~Mayer}{a}
\addAuthor{P.~C.~McDaniel}{f} % check email for ordid number when submitting
% planned feedback by end of weekend
\addAuthor{E.~Niner}{g} % 
\addAuthor{K.~Payne}{f} % approved publishing
\addAuthor{C.~C.~Petitjean}{e}
\addAuthor{R.~Ridgway}{g}
\addAuthor{M.~Russell}{a}
\addAuthor{A.~Sampat}{f}
\addAuthor{J.~Simon}{f}
\addAuthor{I.~D.~Spool}{f}
% approved publishing 
\addAuthor{A.~Tejeda}{k} % Ana Tejeda approved publishing

% Principal investigators for each major lab
\addAuthor{A.~Antognini}{e,j}    % PSI - piE1 muon beam    
% round of edits incorprated 
\addAuthor{K.~R.~Lynch}{g}        % Fermilab
% expects to have comments by end of day 12/22
% expects to have comments 12/23
\addAuthor{S.~O.~Newburg}{a,f}    % NK Labs
\addAuthor{W.~T.~Shmayda}{d}     % Tritium Solutions / LLE
\addAuthor{A.~N.~Knaian}{a,f}     % Last author -Acceleron Fusion 

\date{\today}% It is always \today, today,

% Corresponding author email

\correspondingEmail{ekoukina@acceleron.energy}

% Affiliations
\addAffiliation{a}{Acceleron Fusion, Cambridge, MA, USA}
\addAffiliation{b}{Torion USA, Rochester, NY, USA}
\addAffiliation{c}{Torion Plasma, Barrie, ON, Canada}
\addAffiliation{d}{Tritium Solutions, Rochester, NY, USA}
\addAffiliation{e}{Paul Scherrer Institute (PSI), Villigen, Switzerland}
\addAffiliation{f}{NK Labs, Cambridge, MA, USA}
\addAffiliation{g}{Fermi National Accelerator Laboratory, Batavia, IL, USA}
\addAffiliation{h}{Department of Physics, University of Oxford, Oxford, UK}
\addAffiliation{i}{{Laboratory for Laser Energetics, University of Rochester, Rochester, NY, USA}}
\addAffiliation{j}{Institute for Particle Physics and Astrophysics, ETH Zurich, Switzerland}
\addAffiliation{k}{Massachusetts Institute of Technology, Cambridge, MA, USA}
\addAffiliation{l}{REB Research and Consulting, Oak Park, MI, USA}
\addAffiliation{m}{Department of Physics, Cornell University, Ithaca, NY}

\addKeyword{Tritium}
\addKeyword{Deuterium-Tritium Gas Delivery}
\addKeyword{Muon Catalyzed Fusion}

\titlePage  % <-- this is the magic one

%-------------------------
% Abstract
%-------------------------
\begin{abstract}

We report the design, commissioning, and operation of deuterium--deuterium (DD) and deuterium--tritium (DT) gas delivery systems developed to load a diamond anvil cell (DAC) beam target for muon-catalyzed fusion ($\mu$CF). The DAC approach enables DT fuel to be compressed to GPa pressures at more than twice the liquid density and heated from cryogenic temperatures through $500~\mathrm{K}$, opening access to a substantially expanded parameter range for $\mu$CF kinetics and yield measurements. In this approach, DT is cryo-condensed to a liquid in a minichamber and then compressed in the DAC using a helium-driven pneumatic membrane, achieving high pressures in a millimeter-scale DT sample volume.

A DD gas delivery system was designed and used to validate the experimental apparatus, measure the gas quantities needed for filling, develop operational experience, and collect kinetics and yield data with DD targets. The DT gas delivery system adds tritium-specific capabilities for inventory minimization, secondary containment, and activity monitoring. The DT system integrates depleted-uranium storage beds and a liquid-helium cryogenic condenser used for pressure building and cryopumping. High-purity delivery is provided by a rapid-response palladium permeator. The system is housed in a helium-atmosphere glovebox held at negative pressure with continuous cleanup.

We present the process-and-instrumentation design, a failure-modes-and-effects analysis (FMEA), and data from the experiment’s in-situ Raman spectrometer, which provides direct confirmation of target loading and composition through the optically clear diamond anvils. The 2024 and 2025 DT campaigns achieved repeatable target fills and operation with no measurable tritium releases to the stack, demonstrating safe, high-purity DT loading at novel density--temperature conditions for $\mu$CF studies.

\end{abstract}

%-------------------------
% 1. Introduction
%-------------------------
\section{Introduction}

When a negative muon stops in a mixture of deuterium and tritium, even at ordinary temperatures, it can induce nuclear fusion, releasing a $3.5~\mathrm{MeV}$ alpha particle and a $14.1~\mathrm{MeV}$ neutron \cite{Breunlich1989ANRP, Jones1986Nature}. Most of the time, the muon is released and can catalyze additional fusion reactions. After each fusion, there is an approximately $0.8\%$ probability of the muon sticking to the alpha particle until the muon either decays or is reactivated and returned to the catalysis cycle by collisional processes \cite{Kamimura2023PRC}.

Starting in the 1980s, several research groups observed more than 100 fusion reactions per muon in cold, dense mixtures of deuterium and tritium \cite{Jones1983PRL, Breunlich1987PRL, Ishida1999HI, Bom2005JETP}, raising the possibility that the process could be used for energy production. \cite{Jones1985FT, Ponomarev1990CP, Chaterjee1991IJOP, Eliezer1994FT, Kelly2021JOPE}. A backdrop of renewed interest in fusion more generally, progress in accelerator efficiency \cite{Padamsee2017SST}, muon cooling \cite{Bogomilov2024Nature}, and enhanced computational tools has provided motivation to revisit DT $\mu$CF with modern instrumentation and over a wider range of target conditions.

A critical issue to resolve is that theory and experiment do not fully agree on the kinetics and yield of the process in dense DT mixtures \cite{Rafelski1989PPNP, Petitjean1989FED, Kawamura2004PTPS}. Both theory and extrapolation of experimental trends predict that the yield should increase with increasing temperature, density, and mixture purity \cite{Yamashita2022NSR}. However, the highest yields observed to date are in cryogenic solid DT -- likely because, despite having the lowest temperature, these data points had the highest densities and best mixture purity.

For these reasons, there has been a longstanding desire in the muon physics community to measure the kinetics and yield of muon-catalyzed DT fusion under conditions with simultaneously high temperature, high density, and high gas purity \cite{Jones1987MCF, Petitjean1989FED, Fujiwara2000PRL}. This is what we (the MuFusE collaboration) are currently attempting to do, using a diamond anvil cell (DAC) to compress and heat the fuel.

Gas delivery plays an especially important role in muon-catalyzed fusion experiments because the muon can be efficiently captured and removed from the catalysis cycle by trace chemical impurities \cite{Nagamine2003Book}. A typical muon transfer rate to oxygen is approximately $10^{10}~\mathrm{s^{-1}}$ (at room temperature and 1 liquid hydrogen density (LHD) atomic number density of oxygen)  \cite{Pizzolotto2019ICPP}, compared to the muon decay rate of $4.5\times10^{5}~\mathrm{s^{-1}}$. Thus, even ppb quantities of oxygen (for example) have a measurable impact on the fusion yield.  Gas delivery is also important because fusion rates have been shown to vary with the D/T ratio, the isotopologue fraction (D$_2$ vs DT vs T$_2$), and even with the ortho and para nuclear-spin states of the fuel \cite{Toyoda2003PRL, Adamczak2005PRA}. 

\section{Overview of the Experiment}

\begin{figure}[htbp]
  \centering

  \begin{subfigure}[t]{1\linewidth}
    \centering
    \includegraphics[
      width=\linewidth,
      trim=4cm 4cm 4cm 4cm,
      clip
    ]{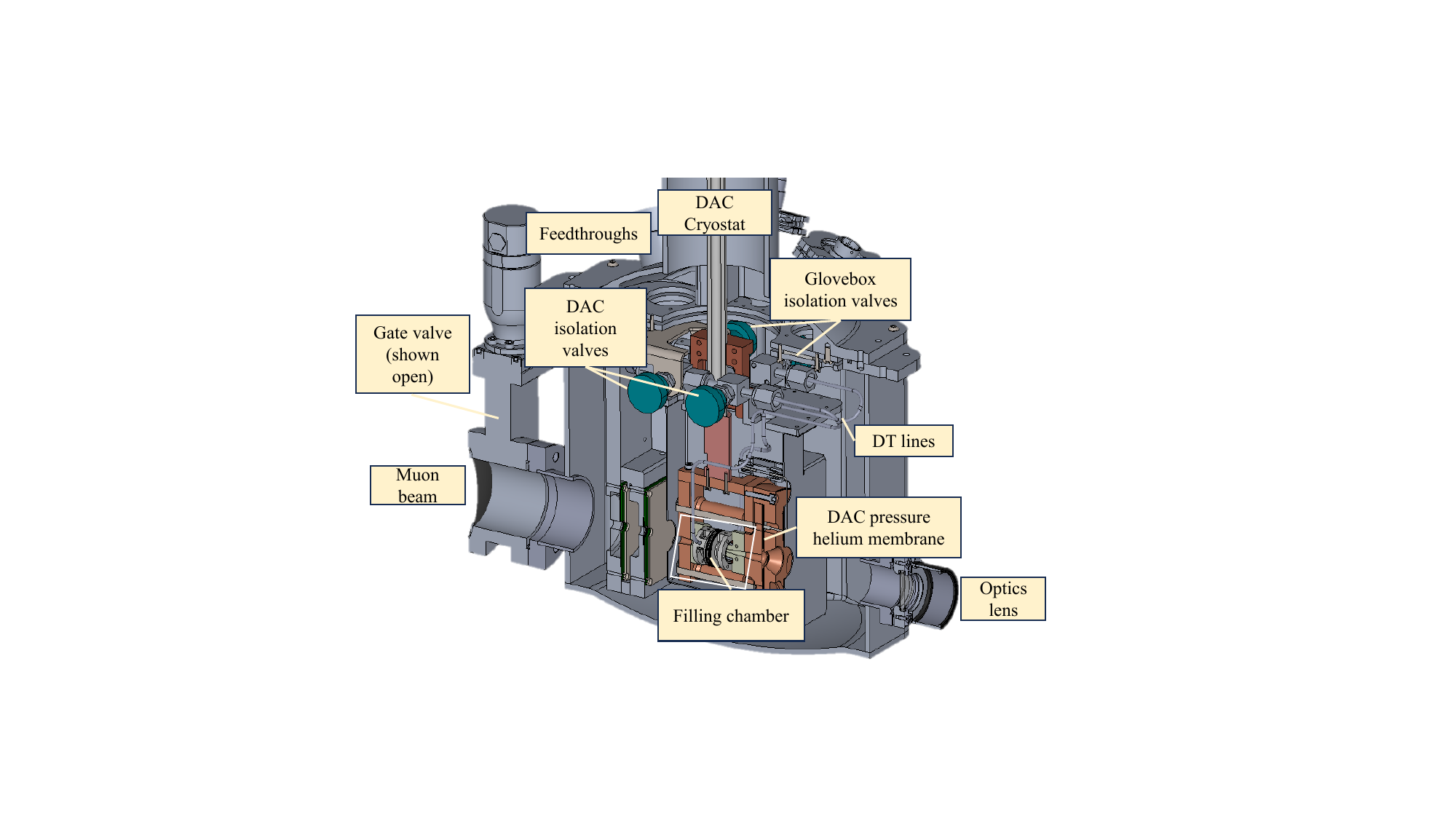}
    \caption{CAD of the diamond anvil cell inside secondary containment connected to DT fill lines through isolation valves. }
  \end{subfigure}
  \hfill
  \begin{subfigure}[t]{1.2\linewidth}
    \centering
    \includegraphics[
      width=\linewidth,
      trim=8cm 4cm 4cm 4cm,
      clip
    ]{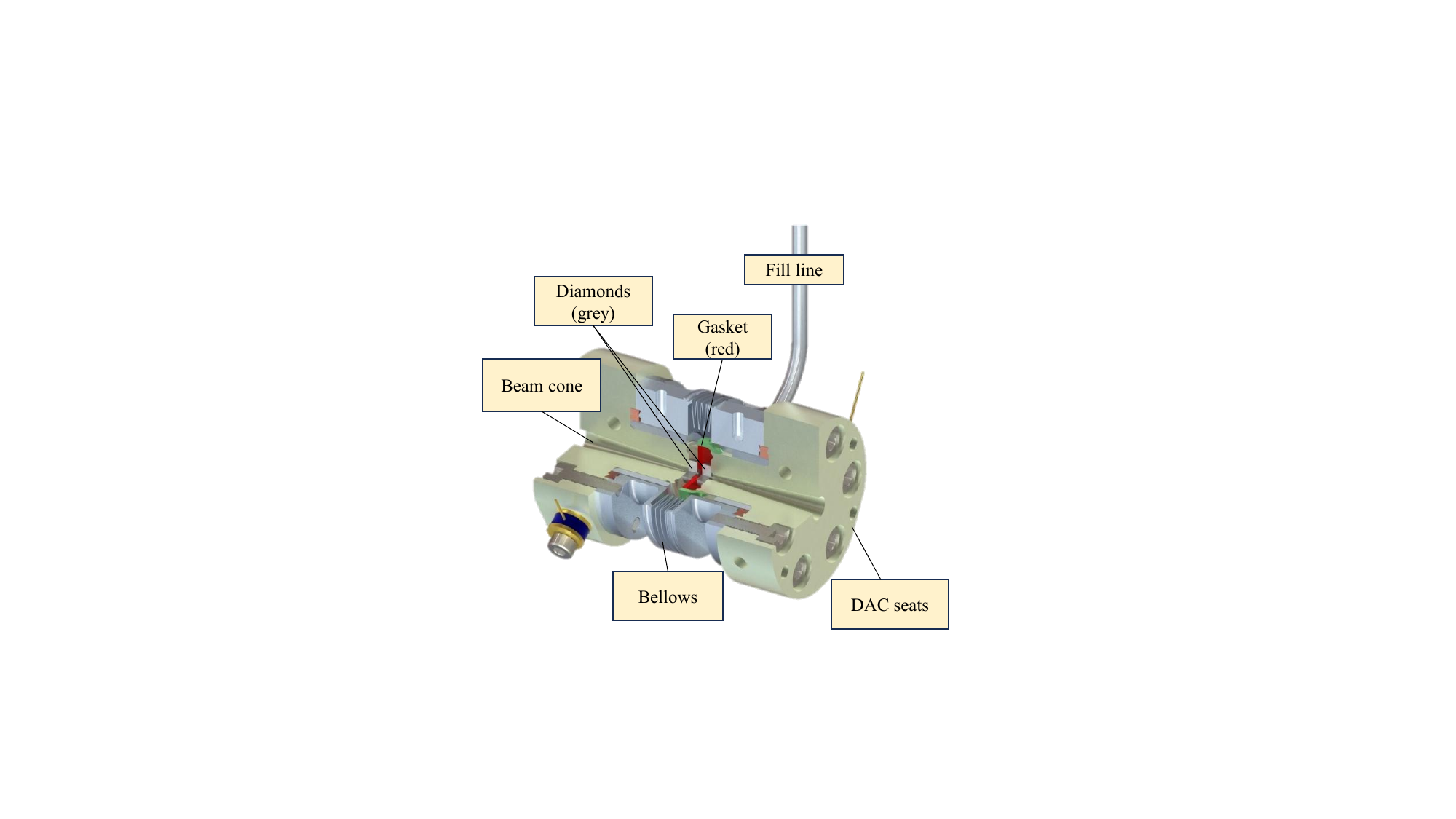}
    \caption{CAD of the diamond anvil cell zoomed in on the filling chamber.}
  \label{fig:minichamber}
  \end{subfigure}

  \caption{CAD of the diamond anvil cell.  Images derived from \cite{KalowInPrep}.}
  \label{fig:DAC}
\end{figure}

The MuFusE collaboration aims to measure the kinetics of muon-catalyzed fusion ($\mu$CF) under temperature and density conditions that extend beyond those previously recorded~\cite{Kawamura2004PTPS,Petitjean1989FED}. To achieve this goal, incoming muons that pass through plastic scintillation counters are stopped in a DAC containing a compressed fuel sample, as shown in Fig.~\ref{fig:DAC}. To facilitate fuel loading, the DAC is surrounded by a minichamber \cite{Silvera1985RSI} which is charged with gas using the gas delivery system. The minichamber is cooled to approximately $20~\mathrm{K}$ by a liquid-helium cryostat, such that about 4 cm$^3$ of target liquid can be condensed inside~\cite{KalowInPrep}. The system, past the permeator that removes impurities from hydrogen, including all seals and joints is constructed from tritium compatible materials \cite{DOETritiumHandbook2015}. A diamond to metal brazing technique was developed to avoid the use of polymers in the construction of the cell \cite{KalowInPrep}.

Once the minichamber is filled with liquid, a helium-filled pneumatic membrane actuator \cite{Zhao2017RSI} closes the DAC, trapping approximately $19~\mathrm{mm}^3$ of condensed fuel which serves as the beam target. This corresponds to a total mass of approximately $4~\mathrm{mg}$ of fuel, and (with a 50/50 D/T ratio) about 24 Ci of tritium.  After the cell is closed, the cell is heated, evacuate the minichamber, and return the remainder of the gas to the delivery system. This system can achieve a temperature range between cryogenic and $500~\mathrm{K}$ using the cryostat and resistive heating, and the target can be compressed to a pressure of 1 GPa using the load supplied by the pneumatic membrane.

The experiments were conducted using the muon beam at the High Intensity Proton Accelerator (HIPA) facility at the Paul Scherrer Institute~\cite{Kiselev2015JRNC}.  The muon beam is tuned to balance stops in the backward and forward anvils, which maximizes stopping in the fuel between the diamonds. A muon stopped in the fuel will generally catalyze reactions until the muon decays or sticks to the produced alpha particle. However, if a muon is captured into the nucleus of a contaminant, fusion will cease prematurely.  Consequently, minimizing contaminants in the gas-loading system and enhancing gas purity is critical to achieving high fusion yields. The products of the reaction are measured using an array of neutron and electron detectors to characterize the fusion rate at different fuel temperatures and pressures achieved within the DAC and the analysis of the physics data is ongoing. The performance of the DAC is reported in ~\cite{KalowInPrep}.

The DAC is swapped out between each run because the gasket in the cell is single-use and requires the DAC to be disassembled to be replaced. To minimize the amount of piping exposed to air during the changes, the DAC is connected through a double valve, allowing the lines to be closed when the DAC is being replaced. The DAC return line is connected to a vacuum pump through double valves as well. All valves are closed when a DAC is replaced. Then the valves are opened starting with the valve closest to the vacuum pump to limit reverse flow of air from the DAC to the gas delivery lines. Once the cell is installed, it is baked out on the system. We run multiple flushes alternating between pressure and vacuum with ultra-pure deuterium while it is warm. The cell is filled with deuterium during cooldown to prepare for a run to avoid leaks from the outside contaminating the cell. 

%-------------------------
% 2. Fueling system
%-------------------------
\section{Gas Delivery Systems}
 
\subsection{Deuterium Gas Delivery System}
\begin{figure}[htbp]
  \centering
  \includegraphics[width=\linewidth,
  trim=1cm 1cm 1cm 0.5cm,
  clip]{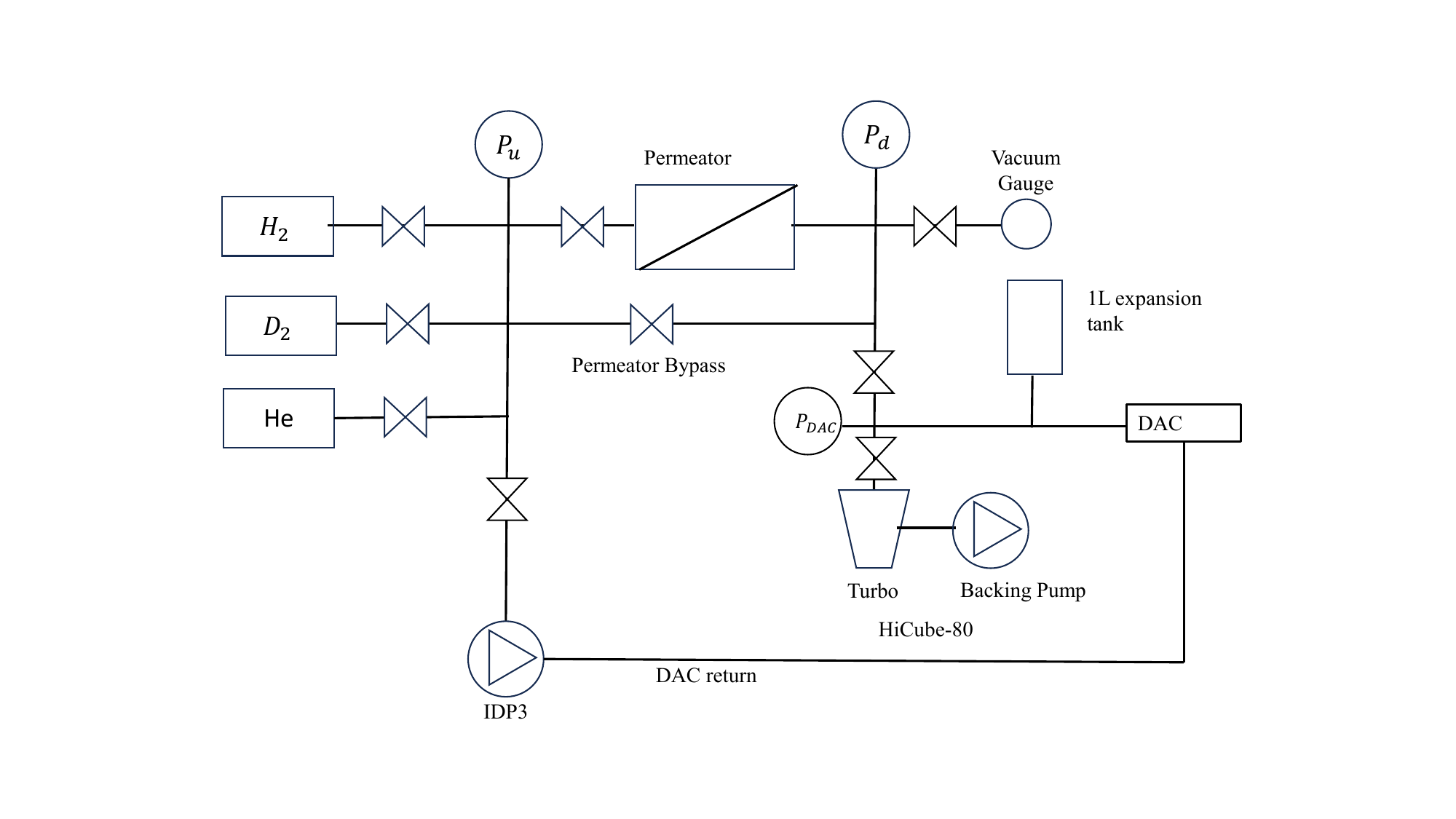}
  \caption{DD gas loading system.}
  \label{fig:DDLoop}
\end{figure}

Figure~\ref{fig:DDLoop} shows the block diagram of the deuterium gas fueling system. The system comprises a hydrogen and deuterium gas supply, hydrogen purification with a palladium-based permeator, vacuum, and a connection to the DAC. The hydrogen and deuterium are passed through the purifier en route to the DAC via a stainless-steel tube. A $1~\mathrm{L}$ expansion tank is provided to ensure that no overpressure event greater than three atmospheres will occur should a liquefied or solidified target have its cooling source fail.

An Agilent IDP-3 dry scroll pump provides the vacuum for an initial evacuation down to $10^{-2}~\mathrm{torr}$. Prior to bakeout, vacuum levels of $10^{-3}$ to $10^{-4}~\mathrm{torr}$ are achieved in the manifold using a HiCube 80 pumping station that consists of a HiPace 80 turbo pump and an MVP 015-2 diaphragm backing pump. Nitrogen is flushed past the vacuum pump exhaust to prevent hydrogen stagnation in the lines going to the stack. 

The manifold, with connections to gas bottles through regulators and vacuum has an output connection to the DAC through about 2 m of 6.35 mm diameter stainless steel tubing in series with about 30 cm of $3.175~\mathrm{mm}$ stainless steel tubing. Welded connections are used where feasible. All non-welded connections use Swagelok stainless steel gasketed VCR connections, apart from industrial pipe fittings from the gas-bottle regulators and Swagelok compression fittings on the upstream side of the hydrogen permeator where introduction of impurities is less critical because they will be removed by the permeator. The components are listed in Table \ref{tab:gas_system}. 

\begin{table}[htbp]
\centering
\caption{Gas handling system instrumentation and components.}
\label{tab:gas_system}
\begin{tabular}{|p{3cm}|p{5cm}|p{6cm}|}
\hline
\textbf{Component} & \textbf{Model} & \textbf{Purpose / Notes} \\
\hline

Upstream pressure gauge ($P_u$)
& MKS 870B-21687 Baratron
& Absolute pressure measurement upstream of the permeator; not bakeable. \\
\hline

Downstream pressure gauges ($P_d$, $P_3$)
& MKS 121A-14364 Absolute Manometer with 121 Signal Conditioner
& Pressure measurement on clean side of permeator; compatible with 150\,\textdegree C bakeout. \\
\hline

Vacuum gauge
& Agilent Varian FRG-720 CF35 Pirani/Bayard-Alpert Combination Gauge
& Vacuum monitoring; bakeable to 150\,\textdegree C with electronics removed. \\
\hline

Upstream valves
& HAM-LET HM20 4VKLC GF4 PCTFE-seat diaphragm valves
& Isolation of gas handling components upstream of permeator. \\
\hline

Downstream valves
& HAM-LET 3LES2C-FV metal-seat diaphragm valves
& Isolation on clean side of permeator. \\
\hline

Primary tubing
& 6.35 mm OD stainless steel tubing
& Approximately 2 m between manifold and DAC. \\
\hline

Secondary tubing
& 3.175 mm OD stainless steel tubing
& Approximately 30 cm immediately upstream of DAC. \\
\hline

\end{tabular}
\end{table}

After a bakeout period of approximately 12 hours, the system is flushed twice with ultra-pure deuterium to displace residual gases. While hydrogen degassing from stainless steel requires higher bakeout temperatures \cite{Redhead1990}, residual hydrogen isotopes are not expected to significantly impact muon-catalyzed fusion performance. The muon transfer rate to hydrogen is small compared to transfer to higher-Z contaminants such as oxygen, which dominate impurity-driven muon losses \cite{Nagamine2003Book,Pizzolotto2019ICPP}.

The system has a bypass path through the FCV4 valve around the hydrogen purifier for loading of additional gases. Use cases for this valve include loading helium for leak checking across valve seats. The deuterium system was used to vet the envisioned operations and to investigate muon-catalyzed DD fusion. The first test operations were performed using argon (MP 84K), which allowed using liquid nitrogen (at 77K) rather than liquid helium (at 4K) as the cryogen. 

The deuterium gas system is controlled by Siemens 1200 series PLCs. One PLC is dedicated to the control of the pneumatic valve states, another controls the 16 zones of bakeout heaters, and a third PLC is used for safety cutoffs in over-temperature conditions.  ROS (Robot Operating System) is used for the PC-based GUI, which is used for real time control of the experiment and data collection. 

The DAC, which contains the minichamber for gas condensation, is cooled by a Lake Shore Cryotronics ST400 cryostat. To safely fill the minichamber, the temperature is managed by a combination of two Lakeshore 335 controllers. One controller is used for temperature control of the cryostat using an internal heater and a type E thermocouple. During cryogenic operations, the temperature of the cryostat is kept below the freezing point of the target gas. The other controller is used to precisely stabilize the minichamber temperature using internal cartridge heaters and cryogenic temperature sensors mounted adjacent to the minichamber. Precise temperature control of the DAC is needed to condense the gas in liquid form. Deposition as a solid could result in an incompletely filled target, and allowing the temperature to range too high would require excessive vapor pressure in the gas system.

%-------------------------
% 2.2 DT Fueling System
%-------------------------
\subsection{Deuterium--Tritium (DT) Gas Delivery System}
During DT runs, the DD system is disconnected from the DAC and the DT system is connected instead using the double valves shown in Fig.~\ref{fig:Glovebox}. The double valves are critical to change the gas system connections without opening long tritiated lines on the DT system between the DAC and the glovebox. 
\subsubsection{Glovebox}

\begin{figure}[htbp]
  \centering
  \begin{subfigure}[t]{\linewidth}
    \centering
    \includegraphics[width=\linewidth]{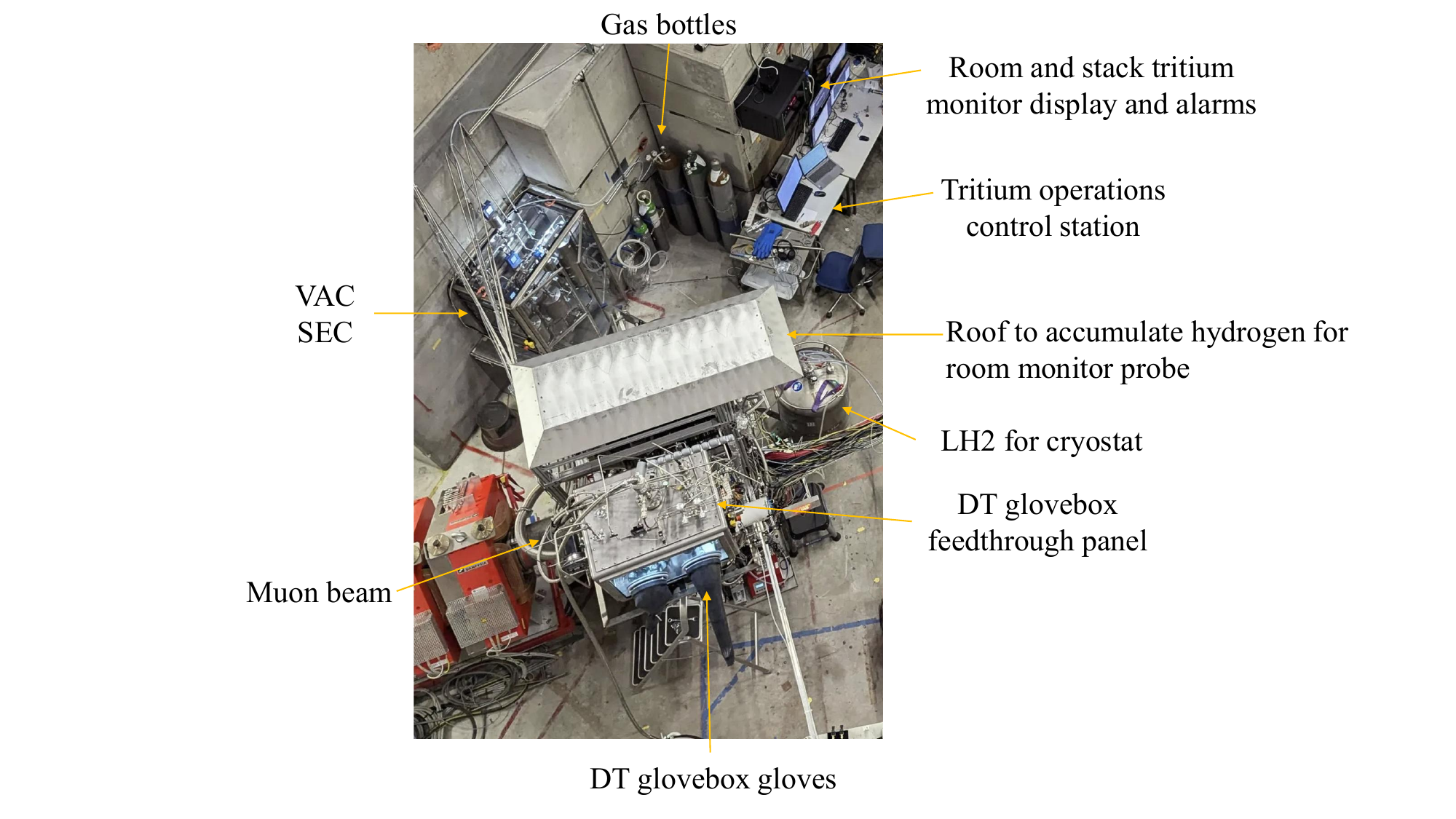}
    \caption{Photo of the experiment at the Paul Scherrer Institute, including the primary loop glovebox.}
    \label{fig:psi_install_photo}
  \end{subfigure}

  \vspace{0.5em}

  \begin{subfigure}[t]{\linewidth}
    \centering
    \includegraphics[width=\linewidth]{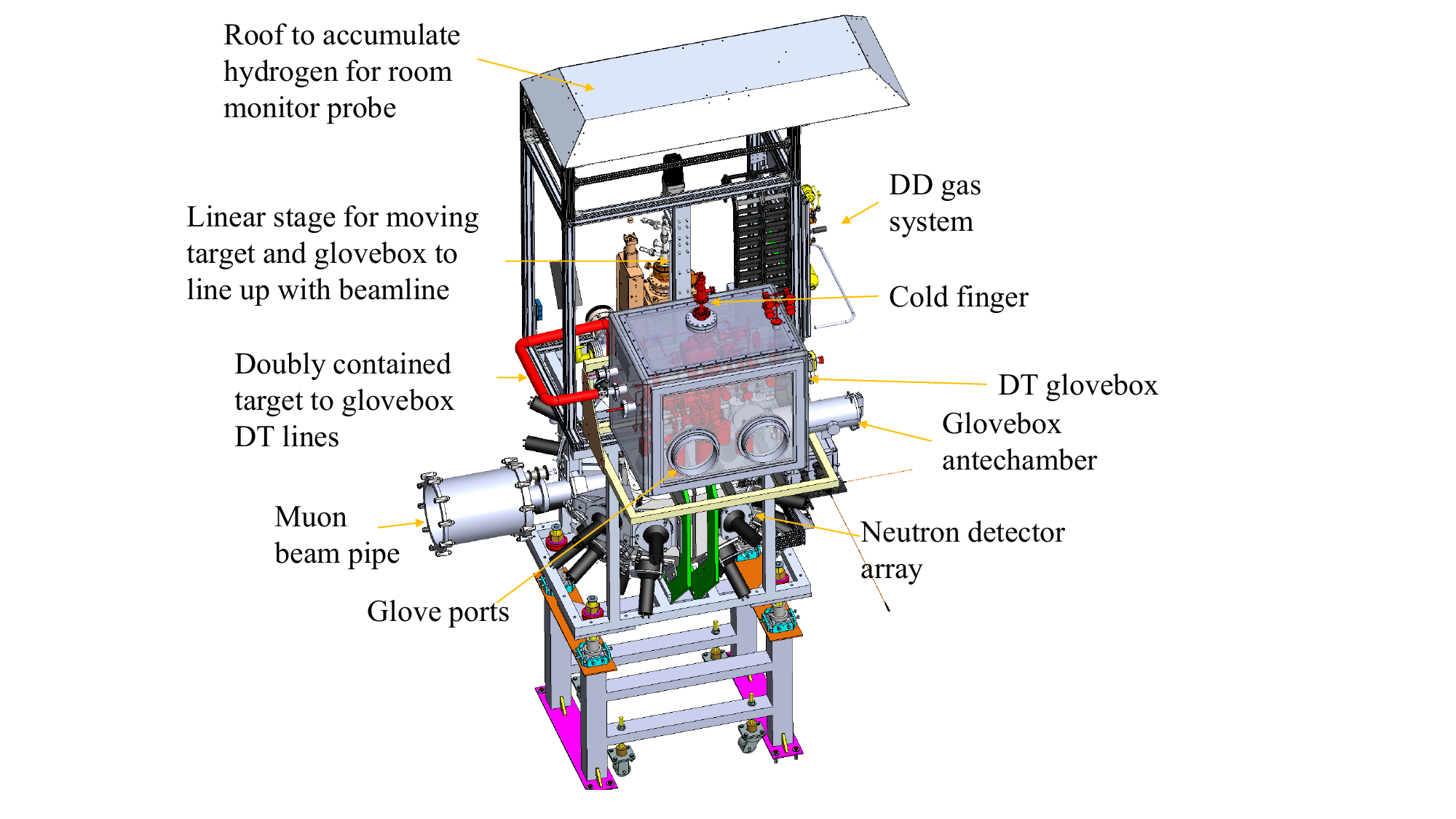}
    \caption{CAD of primary loop glovebox installed at the Paul Scherrer Institute.}
    \label{fig:psi_install_cad}
  \end{subfigure}

  \caption{Primary loop glovebox installation at the Paul Scherrer Institute.}
  \label{fig:Glovebox}
\end{figure}

The DT system is housed inside a hermetically sealed, stainless-steel glovebox held at around 0.2 inches of water negative pressure with respect to ambient conditions. The glovebox is mounted on the vertical stage that moves the target in and out of the beam axis, as shown in Fig.~\ref{fig:Glovebox}. Helium is the working atmosphere within the glovebox. The enclosure helium is cycled through a cleanup system to capture tritium that escapes from the primary process loop in accident scenarios and through small amounts expected with hydrogen permeation through stainless steel ~\cite{Shmayda1992FusionTechnol}.

\subsubsection{DT Loop}

\begin{figure}[htbp]
  \centering
  \includegraphics[width=\linewidth,trim=1cm 3cm 1cm 0.5cm,clip]{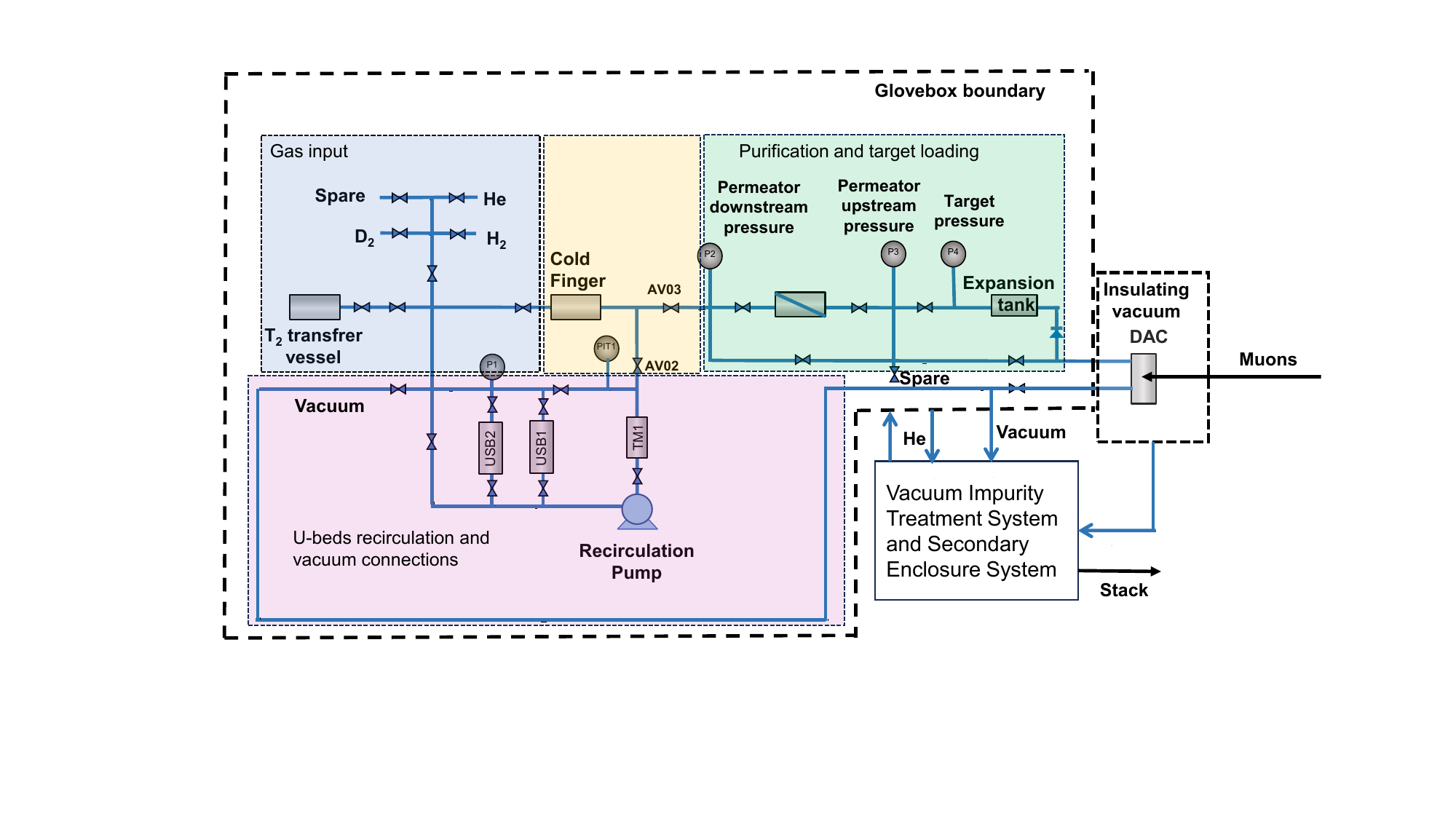}
  \caption{Deuterium--tritium primary loop. Gas input in blue. U-beds recirculation and vacuum connections in pink. Purification and target delivery in green. The cold finger for building pressure is in yellow. USB stands for depleted uranium storage bed and TM: tritium monitor. }
  \label{fig:DTLoop}
\end{figure}

The tritium process system is responsible for making deuterium--tritium mixtures and delivering that gas to the DAC through a hydrogen purifier. The process loop for the DT system, illustrated in Fig.~\ref{fig:DTLoop}, consists of four main functional areas: gas input (in blue), tritium storage (in purple), cold finger (in orange), purification and target delivery (in green). The gas input branch is used for loading tritium into the DT system and adding helium, deuterium, and hydrogen. The tritium storage section consists of depleted-uranium beds for tritium storage, a tritium monitor, and a recirculation pump. The cold finger section consists of a liquid-helium-cooled cryostat that traps DT mixtures to build pressure on the upstream side of the permeator, which removes impurities from the DT fuel. The purification and target loading section contains the connection to the DAC, an expansion tank for safety at cryogenic conditions, and supporting equipment including valves and sensors for loading. The frequently operated valves are pneumatically controlled, while others are manual. The system consists of welded stainless-steel tubing connections and stainless-steel gasket VCR fittings that are helium leak checked to ensure a leak rate of less than $10^{-9}~\mathrm{scc~He/s}$.

The valves upstream of the permeator are Swagelok bellows-sealed BK series pneumatic valves with Vespel stem tips. The valves downstream of the permeator are all-metal HAM-LET 3LSS6CBW pneumatic diaphragm valves.

Charging the test DAC with DT fuel is a multi-step process. DT is desorbed from one of the uranium storage beds (USBs) into the tritium monitor, where the gas is assayed for tritium activity using a 1 L detection volume ionization chamber before transfer to the cold finger~\cite{Shmayda1984ULCM}. The tritium monitor portion of the loop is then isolated, and the gas is desorbed from the cold finger through the permeator to the DAC to remove decay helium and any impurities that may be entrained with the DT. The downstream side of the permeator is actively pumped by holding the DAC at approximately $15~\mathrm{K}$.

Unused DT is returned to the cold USB once the DAC is isolated from the system. Following the experiment, DT within the DAC is also returned to the cold USB, bypassing the permeator. The residual tritium in the loop is circulated over the cold USB to remove as much tritium from the loop as possible. Effluent is evacuated from the process loop to the Vac-SEC system (Vacuum Impurity Treatment and Secondary Enclosure Cleanup System), where trace quantities of tritium are recovered from the effluent before discharge to the stack~\cite{TorionVacITSSEC}.

The USBs are doubly-contained flow-through devices containing 35 g of depleted uranium, designed and built by RC Tritec. Individual secondary containment volumes around each bed capture any tritium that permeates through the walls of the hot primary vessel when the bed is heated to desorb DT and are periodically evacuated.

\subsubsection{Cold Finger}

\begin{figure}[htbp]
  \centering
  % TODO: replace 'fig_cold_finger' with actual filename
  \includegraphics[width=0.9\linewidth]{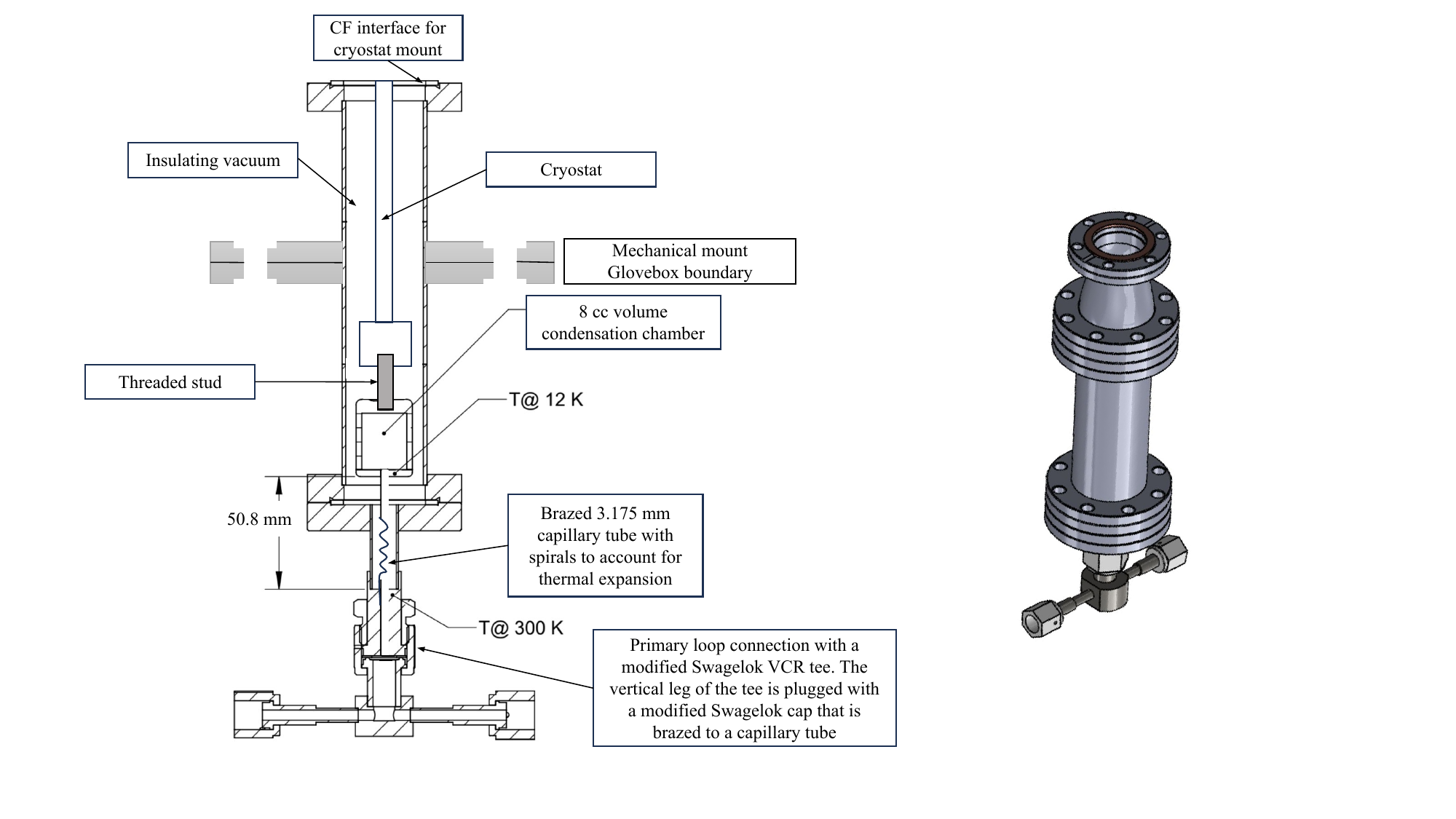}
  \caption{Cross section of the cold finger (left) and CAD model of cold finger assembly (right).}
  \label{fig:ColdFinger}
\end{figure}

The cold finger executes two functions: (1) evacuating DT from volumes by condensing gas on a cryogenic surface and (2) building pressure upstream of the permeator without the need for devices with moving parts such as pumps. The gas delivery systems used in previous muon-catalyzed fusion experiments at the Paul Scherrer Institute, described in \cite{Sherman}, relied on a pump train that included a Normetex pump, two metal bellows pumps, and a highly specialized welded diaphragm pump (DCO) with a discharge pressure of 4 bar. The use of a cold finger to build pressure eliminates the need for tritium compatible pumps. \cite{Sherman}
A cross section of the cold finger is provided in Fig.~\ref{fig:ColdFinger}. The condensation chamber uses a modified VCR tee to couple to the primary process loop via a $3.175$ mm capillary that has been coiled to allow for thermal contraction during cooling. The capillary allows for the cold finger to be at around 4K, while the lines inside the glovebox are at room temperature or higher around the permeator. In operation, the sublimation chamber is between 14K and 24K depending on the amount of time since the introduction of DT into the chamber. The sublimation chamber has a Lakeshore diode coupled directly to it to monitor the temperature, which is used for mapping the gas conditions to the vapor curve. The ARS LT-4B UHV compatible open cycle liquid helium cryostat is mounted to a CF flange that provides an insulative vacuum for the sublimation chamber. The insulating vacuum, also acting as a secondary containment for tritium, is mounted to the glovebox feedthrough panel. This maintains the glovebox atmosphere of positive pressure around the primary loop tee, and provides the insulating vacuum on the outside of the cryostat. This allows the liquid helium flow to be external to the glovebox, while the sublimation chamber is internal to the glovebox.

DT flows from the tritium monitor through the capillary shown in Fig.~\ref{fig:ColdFinger} and condenses within the condensation chamber. Once the transfer is complete, the condensation chamber is warmed to melt the DT ice and pressurize the upstream side of the permeator.

%\begin{figure}[htbp]
  %\centering
  % TODO: replace 'fig_cold_finger' with actual filename
  %\includegraphics[width=0.9\linewidth]{figures/cf.png}
  %\caption{Cold finger thermal performance during sublimation and gas transfer. The upper panel shows the temperature evolution of the system, where Channel 1 corresponds to the sublimation chamber temperature and Channel 2 corresponds to the cold finger temperature. The lower panel shows valve states during operation. AV03 controls the flow path from the sublimation chamber to the tritium monitor, while AV02 controls the flow path from the sublimation chamber to the permeator. Valve states are shown as binary open/closed signals.  }
 % \label{fig:ColdFinger}
%\end{figure}

\subsubsection{Permeator}

The hydrogen purifier is a tritium-compatible, multitube, high-throughput, palladium-silver permeator. The purifier is housed in a secondary container to reduce the heat load within the glovebox and to capture any permeant DT while the permeator primary vessel is at $400^{\circ}\mathrm{C}$. The upstream side is pressurized by the cold finger. The downstream side is pumped by holding the DAC at cryogenic temperatures.

\begin{figure}[htbp]
  \centering
  \includegraphics[width=0.5\linewidth]{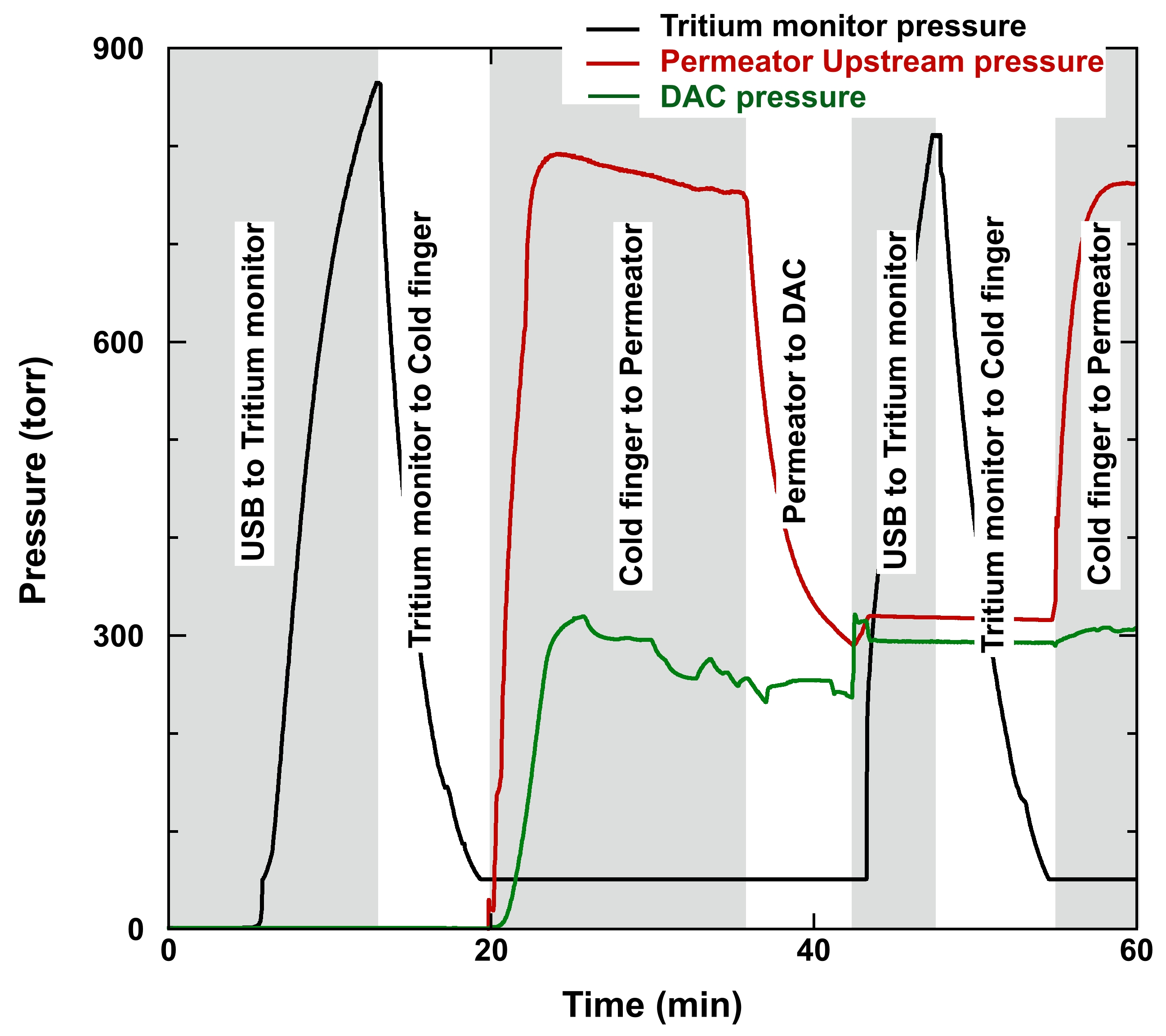}
  \caption{Experimental data from 2024 beamrun: transfer of DT from a uranium storage bed to the DAC through the permeator.}
  \label{fig:PermeatorTransfer}
\end{figure}

An expansion volume is connected to the DAC with a burst disc set to 0.69 bar (10 psi). This is a contingency safety measure should the liquefied or solidified DT fuel in the DAC lose its cooling source. Two sequential DT transfers from a USB through the permeator to the DAC are shown in Fig.~\ref{fig:PermeatorTransfer}. Initially, gas is transferred to the tritium monitor, reaching 860 torr in the first charge and 810 torr in the second charge. Each charge is rapidly condensed on the cold finger then expanded into the volume upstream of the permeator. The upstream gas pressure drops rapidly once the permeator is valved into the loop until the upstream and downstream pressures are approximately equal. The downstream pressure is set by the DT vapor pressure over the DT liquid within the DAC. The DAC is held between 15 and 17 K during this operation.

%-------------------------
% 3. System integration
%-------------------------
\section{System Integration}

The primary process loop, which contains the gas for loading the cell, is surrounded by a secondary enclosure that is equipped with a tritium cleanup system. In the event of a process-loop failure and tritium release from the loop, the gas inside the enclosure is captured as a tritide using the ZrFe getter-bed cleanup system. The cleanup system uses helium cover gas to recover the tritium without converting it to tritiated water.

\begin{figure}[htbp]
  \centering
  % Left: photo
  \begin{subfigure}[b]{0.48\linewidth}
    \centering
    \includegraphics[height=7cm]{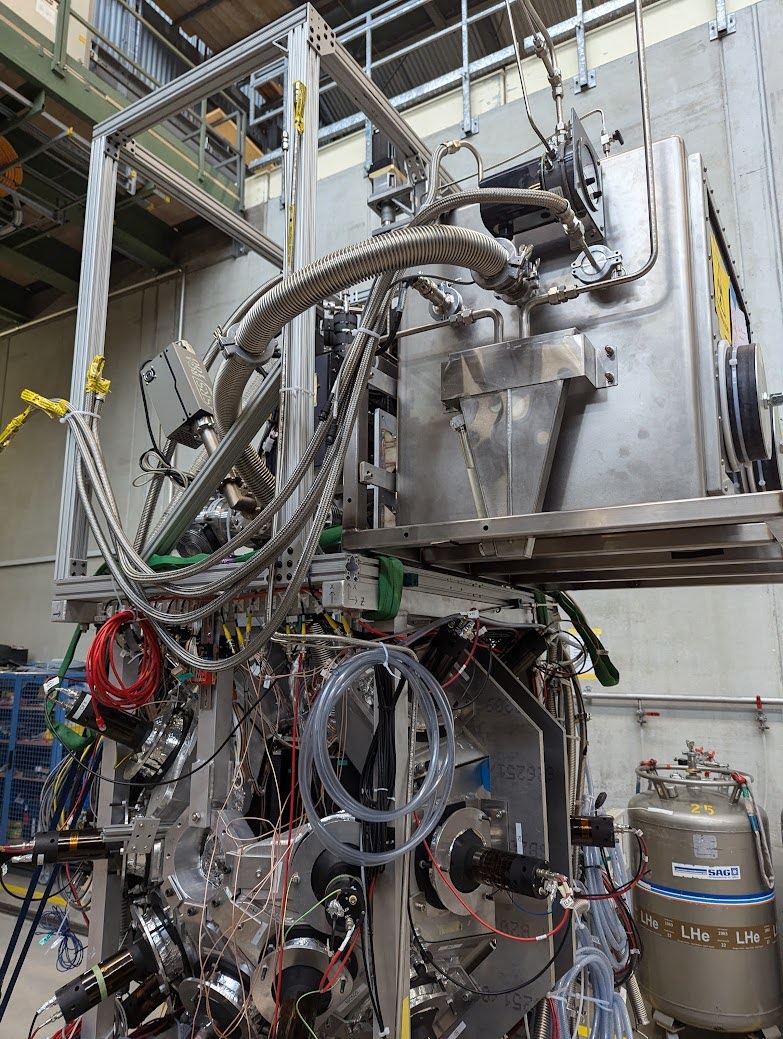} % <-- photo PNG
    \caption{Photograph of the glovebox.}   % subfigure title
    \label{fig:DAC_photo}
  \end{subfigure}
  \hfill
  % Right: CAD
  \begin{subfigure}[b]{0.48\linewidth}
    \centering
    \includegraphics[height=7cm]{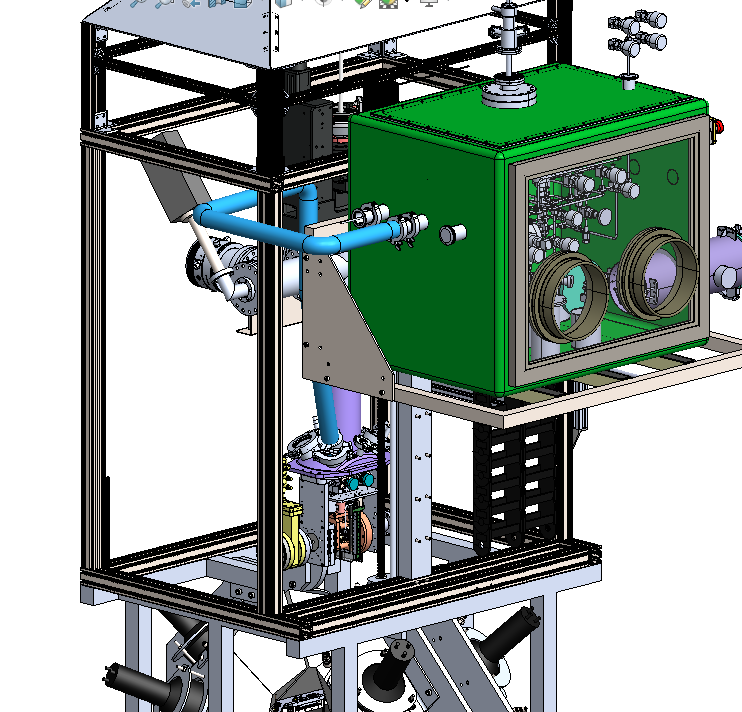}   % <-- CAD PNG
    \caption{CAD model of the glovebox }   % subfigure title
    \label{fig:DAC_CAD}
  \end{subfigure}

  \caption{Glovebox (green) enclosing DT gas system mounted on target system frame. The DT gas line is connected to the target cell via a secondary-enclosure KF50 flexible hose (blue).}
\label{fig:GloveboxToTarget}

\end{figure}

The gas line between the permeator and the target cell is housed in a secondary containment shown as the blue line in Fig.~\ref{fig:GloveboxToTarget}. This volume is part of the secondary containment of the glovebox. Any tritium release into the secondary containment is also collected by the glovebox cleanup system.

The DAC resides in a thermal insulating vacuum chamber. Should a failure of the DAC release tritium into the vacuum chamber, the gas can be recovered to a uranium storage bed. The vacuum chamber is protected by a pressure relief valve. To eliminate air ingress into the vacuum chamber through a leaky pressure relief valve, an MDC Precision 0.4 bar (6 psi) burst disc was installed between the vacuum chamber and the pressure relief valve. The pressure-relief output is connected directly to the stack.

The secondary containment volumes for the USBs and the hydrogen purifier are separated by hand valves to allow evacuation during operation. The DAC return is also connected to vacuum to evacuate any residual gas after DT is returned to the USBs. This effluent is directed to the Vac-SEC system, where tritium is extracted from the effluent before it is sent to the stack. The Vac-SEC system deployed at PSI is a combination of Vac-ITS and SEC systems (Vacuum and Impurity Treatment System and Secondary Enclosure System) designed to detritiate effluents, recover tritium emissions within the glovebox, and maintain glovebox pressure at $20~\mathrm{Pa}$ below ambient conditions within PSI~\cite{TorionVacITSSEC}.

%-------------------------
% 4. DT system commissioning
%-------------------------
\section{DT System Commissioning}

The results of a failure-mode and effects analysis (FMEA) are presented in Appendix~\ref{app:FMEA}. This analysis preceded any deuterium operations. A series of commissioning runs with deuterium was carried out. This exercise included the use of all ancillary equipment, including working within the glovebox and operating the secondary enclosure cleanup system. A fume hood was installed for working with components with trace tritium. 

Tritium gas was delivered in four half-liter vessels, each containing $37~\mathrm{TBq}$ (1000 Ci, 0.10 g) at a pressure of approximately 635 torr (85 kPa). These vessels, which were small enough to be transferred into the helium glovebox through the antechamber, were fixed to the primary process loop one at a time. Tritium was transferred from the gas bottles to the one-liter calibrated-volume tritium monitor using the cold finger in Fig.~\ref{fig:ColdFinger}, assayed to verify the amount of tritium delivered to PSI, and stored on one of the USBs. The transfer vessels were then removed from the glovebox.

Two commissioning runs were performed before proceeding to the full tritium operation: one using a 0.1\% T/D ratio, which put $\sim 1$ Ci at risk, and a second run with a 1\% T/D ratio. Following these two successful runs, the remaining gas was mixed to make a 40\% T/D ratio for the fusion experiments.

%-------------------------
% 5. Results
%-------------------------

\section{Results}

The systems described were successfully deployed in a DT run in the 2024 experimental campaign, followed by three DD and two DT runs in the 2025 campaign. Each run lasted 24--36 hours. Across both campaigns, operations proceeded without any measurable footprint on stack tritium activity at PSI and without dose delivery to personnel.

\begin{figure}[htbp]
  \centering
  \includegraphics[width=0.8\linewidth]{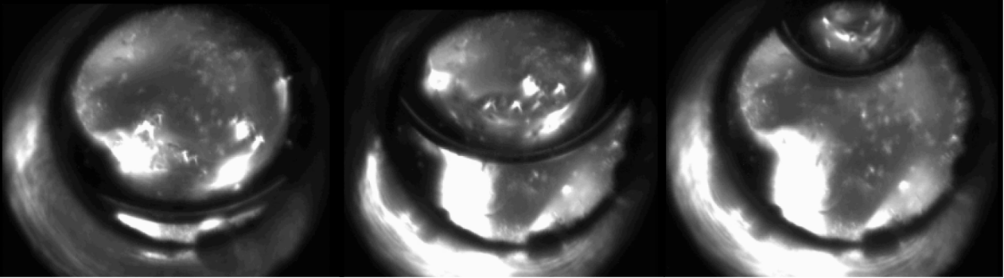}
  \caption{Real time images of the cell used during liquid deuterium/tritium filling. \cite{KalowInPrep}}
  \label{fig:optics}
\end{figure}

The DAC, which has a significantly larger volume than typical DACs, achieved a stable sample volume of $19.2\,\mathrm{mm}^3$ at pressures up to $933$ MPa and temperatures up to 400 K and details on the DAC construction and performance are reported in \cite{KalowInPrep}. Successful gas loading and removal from the DAC were confirmed by direct optical observation of the liquid hydrogen meniscus through the diamond anvils shown in Figure \ref{fig:optics}. Under typical operating conditions, the target fill required approximately 20~minutes after opening the target valve.

\begin{figure}[htbp]
  \centering
  \includegraphics[width=0.8\linewidth]{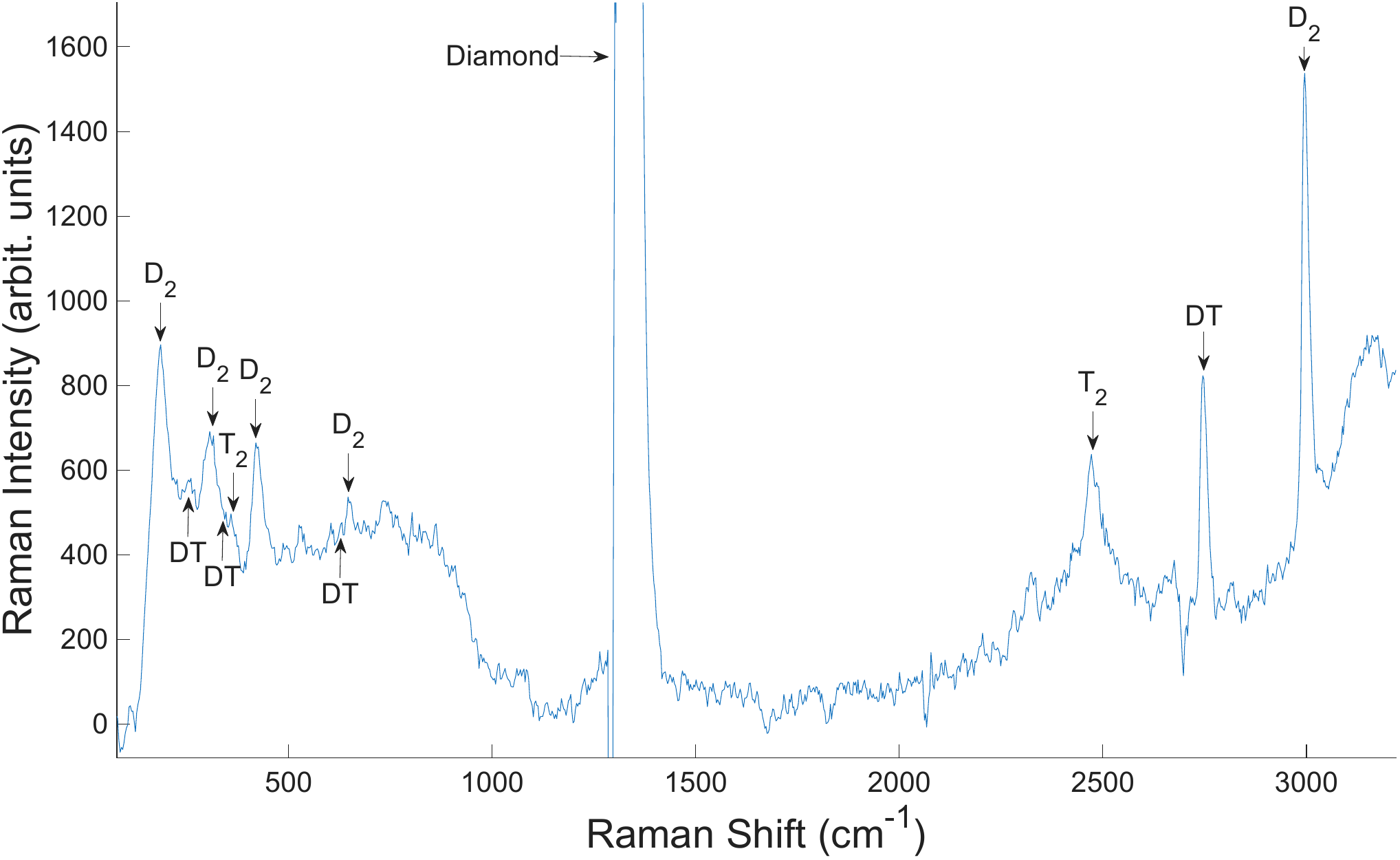}
  \caption{Raman spectra showing prominent hydrogen peaks and a diamond peak from inside the DAC, in the insulating vacuum.}
  \label{fig:raman}
\end{figure}

Target composition was further verified by in-situ Raman spectroscopy of the compressed sample in the anvil cell using 405~nm excitation. The spectrum in Fig.~\ref{fig:raman} is from the second 2025 DT fill. Pre-processing included cosmic spike removal and subtraction of a background spectrum collected minutes earlier with the laser off. Sample spectra were averaged over 7 minutes and background spectra over 10 minutes; the averaged background was then subtracted from the averaged sample. The resulting spectrum shows the rotational and rotation-vibration Raman features of the DT sample in the DAC (Fig.~\ref{fig:raman}).

The muon, neutron, and electron detector data collected show clear evidence of both DD and DT fusion in the anvil-cell target. The data is being analyzed prior to publication and will be covered in a future paper. 

\section{Conclusions}

We have designed, commissioned, and operated DD and DT gas delivery systems that enable safe, repeatable cryogenic fills of a diamond anvil cell target for $\mu$CF experiments. The architecture has demonstrated reliable performance under user-facility beam-run time constraints and has enabled multiple multi-day campaigns with robust tritium confinement.

A key next step toward higher-precision fusion yield measurements is quantitative, on-line verification of trace chemical impurities during fills and after return of gas from the anvil cell. Trace contaminants can reduce the observed fusion yield by capturing muons that would otherwise participate in the catalysis cycle. While our current procedures (e.g. leak checking, bakeout, flushes) are designed to minimize such effects, the present configuration does not provide an on-line assay; therefore, we cannot exclude the possibility of ppb-level gaseous contaminants introduced by outgassing, surface contamination, small leaks, or reactions with interior surfaces. Under tritium operation, beta-stimulated chemistry can provide an additional impurity source~\cite{Shmayda2017FST}.

To address this, we are implementing a hydrogen-elimination mass spectrometry capability on both the DD and DT systems~\cite{Bossard2012HEMS,Meinders1973AC}. In this approach, a palladium-based membrane selectively removes the bulk hydrogen-isotope matrix, enabling sensitive residual-gas analysis of non-hydrogen impurities with a quadrupole mass spectrometer. Combined with the existing in-situ Raman system, this will provide a complete analysis of the chemical, isotopic, isotopologue, and isomeric composition of gas undergoing fusion during each experimental run.  Releasing the gas from between the anvils produces a substantial pressure ($\sim$0.14~bar, $\sim$105~torr) in the gas delivery system, supporting the feasibility of impurity measurements using this approach.

Overall, the DD and DT gas delivery systems presented here constitute a robust architecture for high-purity hydrogen-isotope mixture delivery to small-volume cryogenic beamline targets. Their demonstrated safety performance and operational repeatability are enabling the MuFusE collaboration to collect $\mu$CF kinetics and yield data in a DAC at pressures and densities extending beyond previously explored target conditions.
\section*{Author Contributions}
CRediT authorship contribution statement:

E. Koukina: Conceptualization, Methodology, Investigation, Writing -- original draft, Writing – review \& editing.  
C. Fagan: Conceptualization, Methodology, Investigation, Writing – review \& editing.  
C.~R.~Shmayda: Conceptualization, Methodology, Investigation.  
J.~D.~Kalow: Conceptualization, Methodology, Investigation, Writing - Review \& editing.  
D.~M.~Harrington: Conceptualization, Methodology, Investigation, Writing – review \& editing.
G. Harris: Conceptualization, Methodology, Investigation, Writing – review \& editing.  
K. McCormack: Conceptualization, Methodology, Investigation, Writing -- original draft, Writing – review \& editing.  
M. Mundt: Conceptualization, Methodology, Investigation, Writing -- original draft, Writing – review \& editing.  
K. Lau: Conceptualization, Methodology, Investigation, Software Programming.   
D. Zajac: Conceptualization, Methodology, Investigation, Software Programming. 
M. Koch: Conceptualization, Methodology, Investigation, Writing – review \& editing.
S. Varner: Conceptualization, Methodology, Investigation, Visualization, Writing -- review \& editing.  
A. Golossanov: Conceptualization, Methodology, Investigation.  
S. Bull: Conceptualization, Methodology, Investigation. 
R. Buxbaum: Conceptualization, Methodology. 
W. Stadolnik: Conceptualization, Methodology, Investigation, Writing – review \& editing.

J.~A.~Allen: Software, Conceptualization, Methodology.  
J. Betances: Conceptualization, Methodology, Investigation.
N.~J. Brennan: Conceptualization, Methodology, Investigation, Writing -- review \& editing.  
R. Chaney: Conceptualization, Methodology.
W.~R. Cutler: Conceptualization, Methodology, Investigation, Writing -- review \& editing.
J. Davies: Funding acquisition, Conceptualization, Methodology.
C. Forrest: Conceptualization, Methodology.
P. Gandhi: Conceptualization, Methodology.
J.~T. Hinchen: Conceptualization, Methodology. 
C.~J. Johnstone: Conceptualization, Methodology, Funding acquisition.
K. Kem: Conceptualization, Methodology.
M. Khandaker: Conceptualization, Methodology, Investigation, Software Programming. 
M. Kiburg: Conceptualization, Methodology.
I. Kiniti: Conceptualization, Methodology.
A.~D. Knaian: Investigation
L.~E. Knaian: Conceptualization, Methodology, Investigation.
N.~J.~L.~MacFadden: Conceptualization, Methodology, Investigation, Writing -- review \& editing.
D. Mayer: Conceptualization, Methodology, Investigation.
P.~C. McDaniel: Conceptualization, Methodology, Investigation.
E. Niner:  Conceptualization, Methodology.
K. Payne:  Conceptualization, Methodology, Software Programming. 
C.~C. Petitjean: Conceptualization, Methodology.
R. Ridgway: Conceptualization, Methodology.
M. Russell: Writing -- review \& editing.
A. Sampat:  Conceptualization, Methodology, Investigation, Software Programming. 
J. Simon:  Conceptualization, Methodology.
I.~D. Spool:  Conceptualization, Methodology.
A. Tejeda: Conceptualization, Methodology, Investigation.

\medskip
\noindent\textit{Supervision and leadership:}

A. Antognini: Conceptualization, Methodology, Investigation, Supervision, Project administration, Writing - review \& editing
K.~R. Lynch: Conceptualization, Methodology, Supervision, Project administration, Writing - review \& editing, Funding acquisition.
S.~O. Newburg: Conceptualization, Methodology, Supervision,  Investigation, Project administration, Writing - review \&editing, Funding acquisition.
W.~T. Shmayda: Conceptualization, Methodology, Supervision,  Investigation, Writing-original draft, Writing - review \& editing, Funding acquisition.
A.~N. Knaian: Conceptualization, Methodology, Supervision,  Investigation, Writing-original draft, Writing - review \& editing, Funding acquisition.
%The 14 CRediT roles that can be added are:
%Conceptualization	Ideas
%Data curation
%Formal analysis
% Funding acquisition
% Investigation	
% Methodology
%Project administration	
% Software	Programming
% Supervision	
% Validation
% Visualization	Preparation
% Writing – original draft
% Writing – review & editing

%-------------------------
% 6. Acknowledgments
%-------------------------
\section*{Acknowledgments}
We thank the laboratory directors and scientific advisors who shared their facilities and knowledge with us, including K. Kirch, A. Amato, P. Kammel, M. Hildenbrandt, L. Pedrazzi, S. Harzmann, M. Tisi, P. Meyer, J. Lykken, A.J. Meyer, J. McDaniel, R. Scuzerella, I. Silvera, S. Sinogeikin, C. Hulbert, U. Schroder, D. Newburg, K. Broderick, C. Izzo, V. Glebov and J. Jacobson.  We thank the program management staff at the US Department of Energy, including S. Hsu, A. Diallo, L. Chatterjee, C. Nell, C. Nehrkorn, M. Handley, S. Wurzel, H. Jackson, R. Wineburg, and J. Patel.  We thank investors and advisors, including S. Gorbunov, W. Lau, G. Vlachos, C Kolster, C. Dumas, C. Sacca, E. Helfgott, T. Alberton, K. Anson, S. Bakalar, C. Shapiro, F. Nivi, B. Hall, N. Ravikant, M. Sweeny, M. Chase-Levy, R. Surati, A. Volpe, and N. Heffron. We thank the operations staff at our institutions, including J. Steve, K. Fanning, N. Newburg, and A. Van Loon.  

This work made use of the High Intensity Proton Accelerator (HIPA) and Swiss Research Infrastructure for Particle Physics (CHRISP) at the Paul Scherrer Institute (PSI), the test beam facility (FTBF) at Fermilab, the Laboratory for Laser Energetics (LLE) at the University of Rochester, Industry Lab, the Engine, Torion Plasma, Swagelok Cambridge, Formex, and laboratories at York College and at the Massachusetts Institute of Technology (MIT).   We thank the hundreds of staff members at these institutions and the many job shops we have worked with -- including the crane operators, machinists, welders, accelerator operators, pipefitters, electricians, radiation safety technicians, cryogenic plant staff, cafeteria workers, security guards, stock clerks, administrators, and managers --- who through their skill, dedication, and hard work over the past four years have made this experiment possible.

The information, data, or work presented herein was funded in part by the Advanced Research Projects Agency-Energy (ARPA-E), U.S. Department of Energy, under Award Numbers DE-AR0001271 and DE-AR0001163. The views and opinions of authors expressed herein do not necessarily state or reflect those of the United States Government or any agency thereof.

\section{Disclosure / Use of AI Tools}
ChatGPT (OpenAI) was used only for minor language editing.
\section{Conflict of Interest}
The authors declare that they have no competing financial or non-financial interests.

%-------------------------
% Appendix A: FMEA
%-------------------------
\appendix

\section{Failure-Mode and Effects Analysis (FMEA)}
\label{app:FMEA}

Table~\ref{tab:FMEA} summarizes the FMEA of the DT gas system.

\begin{longtable}{|p{0.35\linewidth}|p{0.55\linewidth}|}
\caption{FMEA of the DT gas system.}
\label{tab:FMEA} \\
\hline
\textbf{Fault condition} & \textbf{Effect} \\
\hline
\endfirsthead

\hline
\textbf{Fault condition} & \textbf{Effect} \\
\hline
\endhead

\hline
\endfoot

\hline
\endlastfoot

Loss of compressed air &
Tritium gas stays in the tubing. All valves fail closed to compartmentalize the tritium. The system can be recovered safely by re-establishing the compressed air supply. \\
\hline

Loss of power &
Tritium gas stays in the tubing. The system can be recovered safely by reconnecting power. \\
\hline

Loss of power or loss of cryogen resulting in temperature increase in the cell that causes overpressure &
Pressure increase is relieved by a rupture disc into a standby vessel. No tritium is released. Tritium is recovered to storage. \\
\hline

Leak inside the vacuum chamber &
Tritium is purged to the secondary enclosure cleanup system, where the tritium is captured. \\
\hline

Leak from plumbing attached to the vacuum chamber &
All tubing has a secondary containment that is filled with helium. Helium purge sends leaking gas to the secondary enclosure cleanup system. \\
\hline

Loss of vacuum &
Tritium in the target heats up and a rupture disc relieves gas to a storage tank. \\
\hline

Connecting deuterium cylinder directly to purge port through operator error &
Gas is sent to the tritium storage and delivery system. Flow rate is limited by a flowmeter on the input. Plumbing is rated to handle the pressure. \\
\hline

Ruptured glove in glovebox &
Vac-SEC senses a jump in the dew point, puts the system in standby, and alerts the operator. \\
\hline

Diamond breaks during compression &
Tritium is released into the vacuum chamber. The operator decides whether to recover the tritium to the tritium storage and delivery system or to pump into the Vac-SEC system. \\
\hline

Leak from minichamber or diamond seals into the vacuum chamber &
Tritium is released into the vacuum chamber and is pumped into the Vac-SEC system to be assayed and stored on getter beds. \\
\hline

Gas membrane rupture releases pressurized helium into the vacuum chamber &
The amount of helium is limited to about 10~cm$^3$ by the solenoid valve prior to the helium regulator, raising the chamber pressure to a fraction of an atmosphere. The helium is pumped into the Vac-SEC system for tritium recovery. \\
\hline

Overpressure in the vacuum chamber &
The vacuum chamber overpressure vent valve releases gas inside the vacuum chamber to the stack until pressure equalizes with atmospheric pressure. \\
\hline

Getter beds in the Vac-SEC overheat &
A backup thermocouple senses bed over-temperature, cuts power to the getter beds, and places the system in standby. \\
\hline

Gas membrane rupture and diamond breaks during compression &
Trace amounts of tritium may contaminate the helium line up to the closed solenoid valve. The helium line is evacuated by the main chamber vacuum pump into the Vac-SEC system. \\
\hline

Uranium storage-bed heater runaway exceeding setpoint temperature &
A watchdog thermocouple cuts power to the heater. \\
\hline

Unexpected release of helium into the glovebox &
The glovebox is protected by a passive system set at a 5~in.\ water column. Helium is directed to the stack. \\
\hline

Overpressurization of the palladium permeator &
Palladium tubes rupture to relieve pressure downstream. No tritium is released into the glovebox, and the experiment is terminated. \\
\hline

Process valve failure &
Operations stop until the errant valve is repaired by working through the gloves in the glovebox. \\
\hline

\end{longtable}

%-------------------------
% References
%-------------------------
\pagebreak
\bibliographystyle{ans_js}
\bibliography{bibliography}

@article{Kawamura2004PTPS,
	title        = {Anomalous Temperature-Dependent Phenomena of Muon Catalyzed Fusion in Solid Deuterium and Tritium Mixtures},
	author       = {Kawamura, N. and others},
	year         = 2004,
	journal      = {Progress of Theoretical Physics Supplement},
	volume       = 154,
	pages        = {233--240},
    url          = {https://doi.org/10.1143/PTPS.154.233}
}

@article{Petitjean1989FED,
	title        = {Muon Catalyzed Fusion},
	author       = {Petitjean, Claude},
	year         = 1989,
	journal      = {Fusion Engineering and Design},
	volume       = 11,
	number       = {1--2},
	pages        = {255--264},
	url          = {https://doi.org/10.1016/0920-3796(89)90023-9},
	issn         = {0920-3796}
}

@article{Redhead1990,
  author  = {P. A. Redhead},
  title   = {Thermal desorption and outgassing of stainless steel},
  journal = {Journal of Vacuum Science \& Technology A},
  volume  = {8},
  number  = {3},
  pages   = {2666--2673},
  year    = {1990},
  url     = {https://doi.org/10.1116/1.576487}
}

@misc{KalowInPrep,
      title={The MuFusE Large-Volume Diamond Anvil Cell for Exploring Muon-Catalyzed Fusion at Higher Pressures and Temperatures}, 
      author={J. D. Kalow and J. T. Hinchen and G. Harris and E. Koukina and D. M. Harrington and P. C. McDaniel and N. J. Brennan and A. Golossanov and I. D. Spool D. Zajac and M. Mundt and S. Varner and M. Russell and S. Bull and K. McCormack and D. Mayer and L. E. Knaian and M. Khandaker and W. Stadolnik and W. R. Cutler and A. Sampat and K. Lau and J. Betances and C. Fagan and C. R. Shmayda and M. Koch and K. Payne and N. J. L. MacFadden and J. Simon and K. Peterson and A. Gami and S. Machavarapu and A. Tejeda and J. Katz and J. A. Allen and R. Chaney and K. Kem and I. Kiniti and E. Garcia Badaracco and K. R. Lynch and P. Gandhi and C. J. Johnstone and E. Niner and C. C. Petitjean and A. Antognini and W. T. Shmayda and S. O. Newburg and A. N. Knaian},
      year={2026},
      eprint={2606.05333},
      archivePrefix={arXiv},
      primaryClass={physics.ins-det},
      url={https://arxiv.org/abs/2606.05333}, 
}

@techreport{Sherman,
  title        = {A Tritium Target System for muCF},
  author       = {J. Zmeskal and P. Ackerbauer and R. H. Sherman and W.B Durham and H.C.Heard and W. Neumann and H. Bossy},
  institution  = {U.S. Department of Energy, Office of Scientific and Technical Information},
  year         = {1990},
  number       = {Proceeding of Muon Calayzed Fusion 1990 Vienna Austria May 28-June 1 1990},
  address      = {Lawrence Livermore National Labratory},
  url          = {https://www.osti.gov/servlets/purl/6242761},
  note         = {OSTI Identifier: 6242761}
}

@article{Shmayda1992FusionTechnol,
	title        = {Inert Gas Secondary Enclosure Clean-Up System},
	author       = {Shmayda, W. T. and Kherani, N. P. and Wallace, B. and Mazza, F.},
	year         = 1992,
	journal      = {Fusion Technology},
	volume       = 21,
	pages        = {616--623},
    url          = {https://doi.org/10.13182/FST92-A29816}
}

@article{Shmayda1984ULCM,
	title        = {Uranium Beds for Temporary Tritium Storage},
	author       = {Shmayda, W. T. and Mayer, P.},
	year         = 1984,
	journal      = {Journal of Less-Common Metals},
	volume       = 104,
	pages        = {239--250},
    url          = {https://doi.org/10.1016/0022-5088(84)90409-0}
}

@misc{TorionVacITSSEC,
  title        = {Vacuum and Impurity System (Vac-ITS) and Secondary Enclosure Cleanup (SEC) System},
  author       = {{Torion Plasma Corporation}},
  howpublished = {Technical documentation, report numbers 6009B-S1-R0 and 6005-S1-02},
  url          = {https://www.torionusa.com/Images_Content/Site1/Files/Products/6009B-S1-R0-Vac-ITS-System-TU.pdf},
  note         = {SEC system PDF: \url{https://www.torionusa.com/Images_Content/Site1/Files/Products/TU-6005-S1-R0-SEC-System.pdf}},
  urldate      = {2025-12-24}
}

@article{Shmayda2017FST,
	title        = {Dependence of Tritium Release from Stainless Steel on Temperature and Water Vapor},
	author       = {Shmayda, W. T. and others},
	year         = 2017,
	journal      = {Fusion Science and Technology},
	volume       = 68,
	pages        = {766--771},
    url = {https://doi.org/10.13182/FST14-913}
}

@article{Breunlich1989ANRP,
	title        = {Muon-Catalyzed Fusion},
	author       = {Breunlich, W H and Kammel, P and Cohen, J S and Leon, M},
	year         = 1989,
	journal      = {Annual Review of Nuclear and Particle Science},
	publisher    = {Annual Reviews},
	volume       = 39,
	number       = {Volume 39,},
	pages        = {311--356},
	url          = {https://doi.org/10.1146/annurev.ns.39.120189.001523},
	issn         = {1545-4134},
	type         = {Journal Article}
}

@article{Kamimura2023PRC,
	title        = {Comprehensive study of muon-catalyzed nuclear reaction processes in the $dt\ensuremath{\mu}$ molecule},
	author       = {Kamimura, M. and Kino, Y. and Yamashita, T.},
	year         = 2023,
	month        = {Mar},
	journal      = {Phys. Rev. C},
	publisher    = {American Physical Society},
	volume       = 107,
	pages        = {034607},
	doi          = {10.1103/PhysRevC.107.034607},
	url          = {https://doi.org/10.1103/PhysRevC.107.034607},
	issue        = 3,
	numpages     = 21
}

@article{Jones1986Nature,
	title        = {Muon-catalysed fusion revisited},
	author       = {Jones, Steven Earl},
	year         = 1986,
	month        = may,
	journal      = {Nature},
	volume       = 321,
	number       = 6066,
	pages        = {127--133},
	doi          = {10.1038/321127a0},
	issn         = {1476-4687},
	url          = {https://doi.org/10.1038/321127a0}
}

@article{Jones1983PRL,
	title        = {Experimental Investigation of Muon-Catalyzed $d\ensuremath{-}t$ Fusion},
	author       = {Jones, S. E. and Anderson, A. N. and Caffrey, A. J. and Walter, J. B. and Watts, K. D. and Bradbury, J. N. and Gram, P. A. M. and Leon, M. and Maltrud, H. R. and Paciotti, M. A.},
	year         = 1983,
	month        = {Nov},
	journal      = {Phys. Rev. Lett.},
	publisher    = {American Physical Society},
	volume       = 51,
	pages        = {1757--1760},
	url          = {https://doi.org/10.1103/PhysRevLett.51.1757},
	issue        = 19,
	numpages     = {0}
}

@article{Breunlich1987PRL,
	title        = {Muon-catalyzed D-T fusion at low temperature},
	author       = {Breunlich, W. H. and Cargnelli, M. and Kammel, P. and Marton, J. and Naegele, N. and Pawlek, P. and Scrinzi, A. and Werner, J. and Zmeskal, J. and Bistirlich, J. and Crowe, K. M. and Justice, M. and Kurck, J. and Petitjean, C. and Sherman, R. H. and Bossy, H. and Daniel, H. and Hartmann, F. J. and Neumann, W. and Schmidt, G.},
	year         = 1987,
	month        = {Jan},
	journal      = {Phys. Rev. Lett.},
	publisher    = {American Physical Society},
	volume       = 58,
	pages        = {329--332},
	url          = {https://doi.org/10.1103/PhysRevLett.58.329},
	issue        = 4,
	numpages     = {0}
}

@article{Bom2005JETP,
	title        = {Experimental investigation of muon-catalyzed dt fusion in wide ranges of D/T mixture conditions},
	author       = {Bom, V. R. and Demin, A. M. and Demin, D. L. and van Eijk, C. W. E. and Faifman, M. P. and Filchenkov, V. V. and Golubkov, A. N. and Grafov, N. N. and Grishechkin, S. K. and Gritsaj, K. I. and Klevtsov, V. G. and Konin, A. D. and Kuryakin, A. V. and Medved', S. V. and Musyaev, R. K. and Perevozchikov, V. V. and Rudenko, A. I. and Sadetsky, S. M. and Vinogradov, Yu. I. and Yukhimchuk, A. A. and Yukhimchuk, S. A. and Zinov, V. G. and Zlatoustovskii, S. V.},
	year         = 2005,
	month        = apr,
	journal      = {Journal of Experimental and Theoretical Physics},
	volume       = 100,
	number       = 4,
	pages        = {663--687},
	doi          = {10.1134/1.1926428},
	issn         = {1090-6509},
	url          = {https://doi.org/10.1134/1.1926428}
}

@article{Ishida1999HI,
	title        = {Measurement of X-rays from muon to alpha sticking and fusion neutrons in solid/liquid D-T mixtures of high tritium concentration},
	author       = {Ishida, K. and Nagamine, K. and Matsuzaki, T. and Nakamura, S. N. and Kawamura, N. and Sakamoto, S. and Iwasaki, M. and Tanase, M. and Kato, M. and Kurosawa, K. and Sugai, H. and Watanabe, I. and Kudo, K. and Takeda, N. and Eaton, G. H.},
	year         = 1999,
	month        = jun,
	journal      = {Hyperfine Interactions},
	volume       = 118,
	number       = 1,
	pages        = {203--208},
	doi          = {10.1023/A:1012669527116},
	issn         = {1572-9540},
	url          = {https://doi.org/10.1023/A:1012669527116}
}

@article{Rafelski1989PPNP,
	title        = {Muon reactivation in muon-catalyzed D-T fusion},
	author       = {H.E. Rafelski and B. Müller and J. Rafelski and D. Trautmann and R.D. Viollier},
	year         = 1989,
	journal      = {Progress in Particle and Nuclear Physics},
	volume       = 22,
	pages        = {279--338},
	url          = {https://doi.org/10.1016/0146-6410(89)90005-7},
	issn         = {0146-6410}
}

@article{Ponomarev1990CP,
	title        = {Muon catalysed fusion},
	author       = {L. I. Ponomarev},
	year         = 1990,
	journal      = {Contemporary Physics},
	publisher    = {Taylor \& Francis},
	volume       = 31,
	number       = 4,
	pages        = {219--245},
	url          = {https://doi.org/10.1080/00107519008222019}
}

@article{Eliezer1994FT,
	title        = {Muon-Catalyzed Fusion — An Energy Production Perspective},
	author       = {Shalom Eliezer and Zohar Henis},
	year         = 1994,
	journal      = {Fusion Technology},
	publisher    = {Taylor \& Francis},
	volume       = 26,
	number       = 1,
	pages        = {46--73},
	url          = {https://doi.org/10.13182/FST94-A30300}
}

@article{Jones1985FT,
	title        = {Engineering Issues in Muon-Catalyzed Fusion},
	author       = {Jones, Steven E.},
	year         = 1985,
	journal      = {Fusion Technology},
	volume       = 8,
	number       = {1P2B},
	pages        = {1511--1521},
	url          = {https://doi.org/10.13182/FST85-A39980}
}

@article{Kelly2021JOPE,
	title        = {An investigation of efficient muon production for use in muon catalyzed fusion},
	author       = {Spencer Kelly, R and Hart, Lucy J F and Rose, Steven J},
	year         = 2021,
	month        = {may},
	journal      = {Journal of Physics: Energy},
	publisher    = {IOP Publishing},
	volume       = 3,
	number       = 3,
	pages        = {035003},
	url          = {https://doi.org/10.1088/2515-7655/abfb4b}
}

@article{Chaterjee1991IJOP,
	title        = {Muon catalysed fusion: the present status},
	author       = {Chaterjee, Lali},
	year         = 1991,
	month        = {may},
	journal      = {Indian Journal of Physics, Part A},
	volume       = 65,
	number       = 3,
	pages        = {175--203},
	issn         = {0252-9262},
	language     = {English},
    url          = {https://files01.core.ac.uk/download/pdf/93520238.pdf}
}

@article{Yamashita2022NSR,
	title        = {Roles of resonant muonic molecule in new kinetics model and muon catalyzed fusion in compressed gas},
	author       = {Yamashita, Takuma and Kino, Yasushi and Okutsu, Kenichi and Okada, Shinji and Sato, Motoyasu},
	year         = 2022,
	journal      = {Scientific Reports},
	volume       = 12,
	number       = 1,
	pages        = 6393,
	issn         = {2045-2322},
	url          = {https://doi.org/10.1038/s41598-022-09487-0},
	date         = {2022-04-16}
}

@article{Fujiwara2000PRL,
	title        = {Resonant formation of d$\mu$t molecules in deuterium: An atomic beam measurement of muon catalyzed dt fusion},
	author       = {Fujiwara, M. C. and Adamczak, A. and Bailey, J. M. and Beer, G. A. and Beveridge, J. L. and Faifman, M. P. and Huber, T. M. and Kammel, P. and Kim, S. K. and Knowles, P. E. and Kunselman, A. R. and Maier, M. and Markushin, V. E. and Marshall, G. M. and Martoff, C. J. and Mason, G. R. and Mulhauser, F. and Olin, A. and Petitjean, C. and Porcelli, T. A. and Wozniak, J. and Zmeskal, J.},
	year         = 2000,
	journal      = {Physical Review Letters},
	volume       = 85,
	number       = 8,
	pages        = {1642--1645},
	url          = {https://doi.org/10.1103/PhysRevLett.85.1642},
	date         = {2000-08-21},
	pmid         = 10970578,
	language     = {eng}
}

@inproceedings{Pizzolotto2019ICPP,
	title        = {FAMU latest results in the measurement of the transfer rate from $\mu$p to oxygen},
	author       = {Pizzolotto, Cecilia},
	year         = 2019,
	booktitle    = {International Conference on Precision Physics and Fundamental Physical Constants-FFK2019},
	volume       = 9,
	pages        = 14,
    url          = {https://doi.org/10.22323/1.353.0013}
}

@article{Toyoda2003PRL,
	title        = {New Insights in Muon-Catalyzed $dd$ Fusion by using Ortho-Para Controlled Solid Deuterium},
	author       = {Toyoda, Akihisa and Ishida, Katsuhiko and Shimomura, Koichiro and Nakamura, Satoshi N. and Matsuda, Yasuyuki and Higemoto, Wataru and Matsuzaki, Teiichiro and Nagamine, Kanetada},
	year         = 2003,
	month        = {Jun},
	journal      = {Phys. Rev. Lett.},
	publisher    = {American Physical Society},
	volume       = 90,
	pages        = 243401,
	url          = {https://doi.org/10.1103/PhysRevLett.90.243401},
	issue        = 24,
	numpages     = 4
}

@article{Adamczak2005PRA,
	title        = {Resonant $d t \mu$ formation in condensed hydrogen isotopes},
	author       = {Adamczak, Andrzej and Faifman, Mark P.},
	year         = 2005,
	month        = nov,
	journal      = {Physical Review A},
	publisher    = {American Physical Society},
	volume       = 72,
	number       = 5,
	pages        = {052501},
	url          = {https://doi.org/10.1103/PhysRevA.72.052501}
}

@book{Nagamine2003Book,
	title        = {Introductory Muon Science},
	author       = {Nagamine, Kanetada},
	year         = 2003,
	publisher    = {Cambridge University Press},
	place        = {Cambridge},
    url          = {https://doi.org/10.1017/CBO9780511470776}
}

@inbook{Jones1987MCF,
	title        = {Can 250+ Fusions Per Muon be Achieved?},
	author       = {Jones, S. E.},
	year         = 1987,
	booktitle    = {Muon-Catalyzed Fusion and Fusion with Polarized Nuclei},
	publisher    = {Springer US},
	address      = {Boston, MA},
	pages        = {73--88},
	isbn         = {978-1-4757-5930-3},
	url          = {https://doi.org/10.1007/978-1-4757-5930-3\_6}
}

@article{Padamsee2017SST,
doi = {10.1088/1361-6668/aa6376},
url = {https://doi.org/10.1088/1361-6668/aa6376},
year = {2017},
month = {apr},
publisher = {IOP Publishing},
volume = {30},
number = {5},
pages = {053003},
author = {Padamsee, Hasan},
title = {50 years of success for SRF accelerators—a review},
journal = {Superconductor Science and Technology}
}

@ARTICLE{Bogomilov2024Nature,
  title     = {Transverse emittance reduction in muon beams by ionization cooling},
  author    = {The MICE Collaboration},
  journal   = {Nat. Phys.},
  publisher = {Springer Science and Business Media LLC},
  month     = {jul},
  year      = {2024},
  url       = {https://doi.org/10.1038/s41567-024-02547-4}
}

@ARTICLE{Silvera1985RSI,
  title     = {Diamond anvil cell and cryostat for low-temperature optical studies},
  author    = {Silvera, Isaac F and Wijngaarden, Rinke J},
  journal   = {Rev. Sci. Instrum.},
  publisher = {AIP Publishing},
  volume    =  {56},
  number    =  {1},
  pages     = {121--124},
  month     =  {jan},
  year      =  {1985},
  url       = {https://doi.org/10.1063/1.1138514}
}

@article{Zhao2017RSI,
    author = {Zhao, J. Y. and Bi, W. and Sinogeikin, S. and Hu, M. Y. and Alp, E. E. and Wang, X. C. and Jin, C. Q. and Lin, J. F.},
    title = {A compact membrane-driven diamond anvil cell and cryostat system for nuclear resonant scattering at high pressure and low temperature},
    journal = {Review of Scientific Instruments},
    volume = {88},
    number = {12},
    pages = {125109},
    year = {2017},
    month = {12},
    issn = {0034-6748},
    url = {https://doi.org/10.1063/1.4999787},
    eprint = {https://pubs.aip.org/aip/rsi/article-pdf/doi/10.1063/1.4999787/13480827/125109_1_online.pdf},
}

@article{Kiselev2015JRNC,
  author    = {Kiselev, Daniela and Baumann, P. and Blau, B. and Geissmann, K. and Laube, D. and Reiss, T. and Sobbia, R. and Strinning, A. and Talanov, V. and Wohlmuther, Michael},
  title     = {The meson target stations and the high power spallation neutron source SINQ at PSI},
  journal   = {Journal of Radioanalytical and Nuclear Chemistry},
  year      = {2015},
  volume    = {305},
  number    = {3},
  pages     = {769--775},
  url       = {https://doi.org/10.1007/s10967-015-3999-3},
  note      = {Published online 28 February 2015}
}

@techreport{DOETritiumHandbook2015,
  title        = {{Tritium Handling and Safe Storage}},
  author       = {{U.S. Department of Energy}},
  institution  = {{U.S. Department of Energy}},
  type         = {DOE Standard},
  number       = {DOE-STD-1129-2015},
  address      = {Washington, DC},
  year         = {2015},
  month        = sep,
  url          = {https://www.standards.doe.gov/standards-documents/1100/1129-AStd-2015/@@images/file},
  note         = {Listed as Sep 16, 2015 on DOE Standards; PDF mirror: https://rampac.energy.gov/docs/default-source/doe-requirements/doe-std-1129-2015.pdf (accessed 2025-12-19)}
}

@techreport{Bossard2012HEMS,
  title       = {New Sensor for Measuring Trace Impurities in Ultra Pure Hydrogen},
  author      = {Bossard, Peter R. and Mettes, Jacques and Breziner, Luis and Gornick, Fred},
  institution = {Power \& Energy, Inc.},
  address     = {Ivyland, PA, USA},
  year        = {2012},
  month       = oct,
  number      = {HEMS\_v6},
  url         = {https://www.powerandenergy.com/pdfs/HEMS_v6.pdf},
  note        = {Version 6 (PDF).},
}

@article{Meinders1973AC,
  author    = {Meinders, Horst},
  title     = {Mass Spectrometric Determination of Impurities in Hydrogen},
  journal   = {Analytical Chemistry},
  year      = {1973},
  volume    = {45},
  number    = {14},
  pages     = {2354--2358},
  publisher = {American Chemical Society},
  url       = {https://doi.org/10.1021/ac60336a009}
}

\end{document}